\documentclass[usegraphicx]{mn2e}
\newcommand{\plotone}[1]{\includegraphics[width=110mm]{#1}} 

\usepackage[fleqn]{amsmath}
\usepackage{amssymb}
\usepackage{multirow}
\usepackage{comment}
\usepackage{epsfig}

\graphicspath{{./}{figures/}}

\newcommand{\e}[1]{\ensuremath{{}_{\text{#1}}}}
\newcommand{\U}[1]{\ensuremath{\mathrm{~#1}}}  
\newcommand{\unit}[1]{\U{#1}}
\newcommand{\myfm}[1]{\mbox{$#1$}}
\newcommand{\solar}{\myfm{_\odot}}
\newcommand{\solarm}{\unit{M\solar}}

\newcommand{\f}[2]{{\ensuremath{
        \mathchoice
        {\dfrac{#1}{#2}}
    {\dfrac{#1}{#2}}
        {\frac{#1}{#2}}
        {\frac{#1}{#2}}
}}}
\newcommand{\tf}[2]{\ensuremath{#1/#2}}
\newcommand{\paf}[2]{\ensuremath{\left(\f{#1}{#2}\right)}}
\newcommand{\pa}[1]{\ensuremath{\left(#1\right)}}
\newcommand{\pac}[1]{\ensuremath{\left[#1\right]}}
\newcommand{\moyenne}[1]{\ensuremath{\left\langle#1\right\rangle}}
\newcommand{\qetq}{\quad\text{and}\quad\null}
\newcommand{\dd}{\ensuremath{\mathrm{d}}}
\newcommand{\Int}[2]{\int_{#1}^{#2}}

\usepackage{calc}

\newcommand{\tableline}{\hline}
\newcommand{\acknowledgments}{\section*{Acknowledgments}}

\makeatletter
\let\jnl@style=\rmfamily
\def\ref@jnl#1{{\jnl@style#1}}
\newcommand\aj{\ref@jnl{AJ}}
\newcommand\apj{\ref@jnl{ApJ}}
\newcommand\apjl{\ref@jnl{ApJ}}
\newcommand\apjs{\ref@jnl{ApJS}}
\newcommand\aap{\ref@jnl{A\&A}}
\newcommand\aapr{\ref@jnl{A\&A~Rev.}}
\newcommand\aaps{\ref@jnl{A\&AS}}
\newcommand\baas{\ref@jnl{BAAS}}
\newcommand\mnras{\ref@jnl{MNRAS}}
\newcommand\pasp{\ref@jnl{PASP}}
\newcommand\pasj{\ref@jnl{PASJ}}
\newcommand\nat{\ref@jnl{Nature}}
\newcommand\aplett{\ref@jnl{Astrophys.~Lett.}}

\makeatother

\newlength{\DiamondWidth}
\newlength{\cdotWidth}
\newlength{\centeringLength}
\newcommand{\DiamondJJ}{
		\hbox{\Large$\diamond$}\hskip-\DiamondWidth
		\hskip\centeringLength\hbox{\Large$\cdot$}
		\hskip\centeringLength
}

\title[Mass of Dense Star Clusters in Starburst Galaxies]{On the Mass of Dense
Star Clusters in Starburst Galaxies from Spectro-Photometry}

\author[J.-J. Fleck, C.~M. Boily, A. Lan\c{c}on and S. Deiters]{J.-J. 
Fleck$^1$, C.~M.  Boily$^1$, A. Lan\c{c}on$^1$ and S. Deiters$^2$ \\
$^1$Observatoire astronomique, 11 rue de l'Universit\'e, F-67000
Strasbourg, France \\
$^2$School of Mathematics, University of Edinburgh, King's Buildings, 
Edinburgh EH9 3JZ, Scotland, UK}

\date{Accepted 2006 March 24.  Received 2006 March 24; in original form 2006 
February 23}

\pagerange{\pageref{firstpage}--\pageref{lastpage}} \pubyear{2006}

\begin{document}

\maketitle

\label{firstpage}

 \begin{abstract}

The mass of unresolved young star clusters derived from spectro-photometric
data may well be off by a factor of 2 or more once the migration of massive
stars driven by mass segregation 
 is
accounted for. We quantify this effect for a large set of cluster parameters,
including variations in the stellar IMF, the intrinsic cluster mass, and mean
mass density. Gas-dynamical models coupled with the Cambridge stellar
evolution tracks allow us to derive a scheme to recover the {\it real}
cluster mass given measured half-light radius, one-dimensional velocity
dispersion and age. We monitor the evolution with time of the ratio of
real to apparent mass through the parameter~$\eta$. 
 When we compute $\eta$ for rich star clusters, we find non-monotonic
evolution in time when the IMF stretches beyond a critical cutoff mass
of $25.5\solarm$.
 We also monitor the rise of color gradients between the inner and outer
volume of clusters: we find trends in time of the stellar IMF power indices
overlapping well with those derived for the LMC cluster NGC 1818 at an age of
$30\U{Myr}$. We argue that the core region of massive Antenn\ae{} clusters
should have suffered from much segregation despite their low ages.  We apply
these results to a cluster mass function, and find that the peak of the mass
distribution would appear to observers shifted to {\it lower} masses by as
much as $0.2$ dex. The star formation rate (SFR) derived for the cluster 
population is then underestimated by from $20$ to $50$ per cent.

\end{abstract}

\begin{keywords}
methods: numerical -- stars: evolution -- stars: kinematics --
stars: luminosity function, mass function -- galaxies: star clusters --
galaxies: clusters: individual: NGC 1818 
\end{keywords}

\section{Introduction}

Star clusters are traditionally thought of as old primordial structures with
ages ranging up to a Hubble time, hence the emphasis of theoretical modeling
on their long-term evolution (e.g. Spitzer 1987; see Meylan and Heggie 1997
for a review). However the wealth of massive young clusters with spectroscopic
ages of less than $100\U{Myr}$ observed with the HST 
 in interacting galaxies (e.g. the Antenn\ae{}, M81/82) has driven much
interest in recent years to the understanding of their formation and early
evolution.
 Closer to us, the Large Magellanic Cloud hosts a set of young clusters (Elson
et al. 1989; Elson 1991), of which some show signs of primordial mass
segregation.
 The cluster NGC 1818 is one such cluster where colour gradients are difficult
to account for other than as a result of their formation history (Hunter et
al. 1997). 
 These issues clearly have bearing on the subsequent dynamical evolution and
photometric properties of clusters and their immediate surroundings (e.g.,
Lamers et al. 2006).

 To understand very young clusters in quantitative detail poses a particular
challenge to theorists since realistic models must account both for the
dynamics and the rapidly-evolving photometric properties of a young stellar
population. 
 In this contribution, we aim to identify evolutionary trends that lead
potentially to large errors when measuring the mass of an unresolved 
cluster with photometry and spectroscopy, an effect
that bears on all observable cluster properties. 

When in dynamical equilibrium,
 the virial theorem gives an exact relation between mass $M$ and mean
three-dimensional velocity dispersion $\sigma$:
 \begin{equation} 
	M = \frac{|W|}{\sigma^2} \equiv \frac{r\e{g}\,\sigma^2}{G}, 
		\label{virial:theorem}
 \end{equation} 
 where $W$ is the gravitational potential energy, $G$~the gravitational
constant and $r\e{g}$ a radius so defined.  All quantities
entering Eq.~\eqref{virial:theorem} must be matched with observables in
projection.  The line-of-sight velocity dispersion $\sigma\e{los}^{\ 2}$
equals $\sigma^2/3$ for an isotropic velocity field, while the gravitational
radius may be expressed in terms of the projected half-light radius
$R\e{hl}$\footnote{By convention and when possible, projected quantities are
denoted with upper case letters.} as
 \begin{equation} 
	r\e{g} \approx \f{5}{2} \times \f{4}{3} \, R\e{hl} ,\label{rg:Rhl}
 \end{equation}
 where the numerical factor $5/2$ gives a rough conversion to a wide-range of
clusters fitted with a King mass profile, and the factor $4/3$ comes from
projection on the sky (e.g., McCrady et al. 2003; Spitzer 1987 \S1.2). 
Equation~(\ref{rg:Rhl}) applies when light traces mass throughout the cluster.
With this in mind we may isolate for M in Eq.~\eqref{virial:theorem} to 
obtain
 \begin{equation} 
	M = \eta \frac{R\e{hl}\, \sigma\e{los}^{\ 2}}{G},\label{virial:observation}
 \end{equation}
 where the dimensionless parameter $\eta \simeq 10$. Several authors have used
$\eta \approx 10$ combined with spectro-photometric data to derive $M$
from Eq.~\eqref{virial:observation}. Such mass estimates can be compared to 
masses
derived from synthetic stellar populations of the same King models fitted to
the data to set constraints on the stellar IMF.  For instance, the stellar
population of massive Antenn\ae\ clusters appears to be inconsistent with a
universal (field) stellar IMF (Mengel et al. 2002; Smith \& Gallagher 2001). 
And several clusters in the galaxy M82 are found to be over-luminous with
respect to their mass, suggesting a top-heavy stellar IMF in these clusters
(Smith \& Gallagher 2001; McCrady et al. 2003).

The above studies have taken a fixed value of $\eta$ for clusters of ages up
to $t = 100\U{Myr}$. This simplification, while intuitively appealing, was 
shown
recently not to be of universal use for
 massive clusters (Boily et al.  2005). Dense, populous clusters will fill the
entire range of stellar masses drawn from the IMF. This has the effect of
dramatically reducing the mass-segregation time-scale compared
with the 
relaxation time-scale and driving heavy
(bright) stars to the centre of the cluster. The measurements of $R\e{hl}$ and
$\sigma\e{los}$ are then biased to values associated with a specific stellar
population, and not the cluster as a whole as assumed in 
deriving Eq.~\eqref{virial:observation}. 
  When the density of the cluster is low (at a given number of stars), this
bias is reduced and $\eta$ remains constant over time to a good approximation.
Using theoretical gas-dynamical models, Boily et al. (2005) found a rough
threshold of mean surface density such that when the initial cluster density
is $\langle\Sigma\rangle \approx 10^4\U{M\solar.pc^{-2}}$ or
 more, masses derived assuming constant $\eta$ systematically underestimate
the real mass by a factor of a few.

This contribution explores a fuller range of parameters and addresses other
issues (colour gradients, systematics) not covered by Boily et al. (2005). In
the next section we briefly recall the dynamical time-scales relevant to the
problem and show explicitly why a bias should be anticipated when deriving the
mass of rich, dense clusters. In~\S3, we give details of the numerical
approach used to conduct the study. In \S4, we discuss how the models were
analysed and quantify numerical and systematic errors. 
 \S5 presents the results of our survey. 
 In \S6 we apply these results to the profiling of the stellar mass function
and colour gradients in clusters. We also explore their implication for a
cluster mass function, and show that the star formation rate inferred from
cluster populations may be strongly biased to lower values. The concluding
section introduces a diagram that relates observed cluster properties to their
underlying potential and draws attention to future developments.

\section{The dynamics of mass segregation}

Two conditions have to be met for Eq.~(\ref{virial:observation}) to be
applicable. 
 First, all stellar components should be in dynamical equilibrium, a sensible
assumption whenever the cluster age exceeds the virialisation time-scale, 
i.e.,
several system crossing time $t\e{cr}$, where
 \begin{equation} 
	t\e{cr} \equiv 2\,r\e{hm}/\sigma \label{crossing:time}
 \end{equation}
 with $r\e{hm}$ being the spherical half-mass radius. 
 The mechanics of virialisation leads to equilibrium velocity distribution
functions independent of stellar masses when all stars have the same radial 
distribution. Collisional gravitational dynamics,
on the other hand, sets a trend towards equipartition of kinetic energy
between stars of different masses as the system evolves.
 The resulting instability has been studied by Spitzer (1969), see also 
Khalisi et al. (2006) for a recent work.
 For two-component
systems of individual masses $m_1$ and $m_2$, 
this situation is expressed as
 \begin{equation}
	\f{1}{2}\,m_1\,{\sigma_1}^2 = \f{1}{2}\,m_2\,{\sigma_2}^2
	\label{kinetic:equipartition}
 \end{equation}
 and hence the ratio of squared velocities of the stars equals the inverse
ratio of their masses: heavier stars have lower velocities on the mean, and
drop to the centre of the cluster. 
 The state of dynamical equilibrium is a
good approximation to the dynamics only when the migration of the heavy stars
takes place over long time-scales.

Secondly, the light should trace the mass so that half-light and half-mass
radii are identical.
 When this is not the case, the mass $M$ may be derived from either of the two 
relations
 \begin{equation}M =\eta_0\,\f{R\e{hm}\,\sigma\e{m1d}^{\ 2}}{G} \qetq
   M =\eta  \,\f{R\e{hl}\,\sigma\e{los}^{\ 2}}{G}\end{equation}
where $\sigma\e{m1d}$ is the mass-weighted velocity dispersion in projection
along the line of sight, and $\sigma\e{los}$ its light-weighted analog, i.e. 
the line of sight velocity dispersion most directly accessible to observation.
 $\eta_0 \approx 10$ is the reference value mentioned already. These relations
combine to give
 \begin{equation}
	\eta 
=\eta_0\,\f{R\e{hm}\,\sigma\e{m1d}^{\ 2}}{R\e{hl}\,\sigma\e{los}^{\ 2}}
		\label{eta:eta0}
 \end{equation}
 and hence $\eta \ne \eta_0$ whenever light and mass follow different runs
with $R$. Since bright stars carry all the light but a small fraction of the
total mass, we expect $\eta > \eta_0$ as the massive stars migrate to the
centre, and both $R\e{hm}$ and $\sigma\e{m1d}$, weighted through the
near-static total mass distribution, remain essentially constant.

\subsection{Characteristic time-scales}

 The total mass distribution of a star cluster evolves slowly over
 the relaxation time $t\e{r}$ of single-population clusters given by 
 \begin{equation}
	\f{t\e{r}}{t\e{cr}} \simeq \f{0.138}{2}
	\pac{\f{r\e{hm}}{r\e{g}}}^{\tf{1}{2}}\,\f{N}{\ln(0.4\,N)}
	\label{relaxation:time}
 \end{equation}
 which
 is identical to Meylan \& Heggie (1997, \S7) once 
Eq.~\eqref{crossing:time} is taken into account 
 where we used $2r\e{hm}$ in the definition rather than $r\e{g}$.
 $N$ is the number of member stars and
 the ratio $\tf{r\e{hm}}{r\e{g}} \approx 0.4$ for a wide range of model fits
to observed clusters. 
 
 With $N = 500\,000$, $r\e{hm}=\tf{4}{3}\,R\e{hm}=1.3\U{pc}$ (suggested from 
massive clusters data) and 
${\sigma}=\sqrt{3}\,\sigma\e{los}=26\U{km.s^{-1}}$, we find a relaxation time 
\begin{equation}t\e{r} \approx 1800\,t\e{cr} \approx 180\U{Myr} \end{equation}
 and hence no massive cluster with an age of less than $100\U{Myr}$ would be
expected to show signs of evolution due to two-body relaxation. It is this
argument that led to the widely used assumption of no evolution of clusters in
young starburst galaxies. However, Farouki \& Salpeter (1982) pointed out that
the trend toward equipartition is accelerated as the mass spectrum $\{m_j\}$
of stars is widened; their analysis suggests that the mass segregation will
take place on a time-scale $t\e{ms}$ given by (Spitzer 1987)
 \begin{equation}
	\f{t\e{ms}}{t\e{r}} \simeq \f{\pi}{3}\,\f{\moyenne{m}}{m\e{max}}
	\f{\overline{\rho}}{\rho}\pac{\f{r\e{hm}}{r\e{g}}}^{\tf{3}{2}},
\label{segregation:time}
 \end{equation}
 where $\overline{\rho}\equiv (M/2)/(4\pi\, r\e{hm}^{\ 3}/3)$ is the mean
density inside the three-dimensional half-mass radius (an over-line denotes
averaging over space, and brackets averaging by mass); and $m\e{max} =
\max\{m_j\},\ j = 1 .. J$.  Note that $t\e{ms}\approx t\e{r}$ when the mass
spectrum is narrow, i.e., we recover the single-component cluster relaxation
time. 
 The mean mass \moyenne{m} $\approx 0.7\solarm$ for a standard Kroupa IMF
(Kroupa 2002).
 Going back to the numerical example given in the above, setting
$m\e{max}=20\solarm$ in Eq.~\eqref{segregation:time} already reduces the mass 
segregation time-scale $t\e{ms}$
to a few Myr, suggesting that mass segregation will be effective over the
life of massive stars. 

\subsection{Example: 2-mass-component systems}\label{M:to:L}

Consider a cluster with two stellar masses, 
$m_1 < m_2$,  mean surface density $\Sigma$ and mass weighted
 squared velocity 
dispersion $\moyenne{\sigma^2}$. Each component $i$ has a velocity 
dispersion $\sigma_i$, a surface density $\Sigma_i$ and a surface 
brightness $\Lambda_i$.
The equilibrium  velocity dispersion assuming 
Eq.~\eqref{kinetic:equipartition} 
leads to
 \begin{eqnarray}
 	\moyenne{\sigma^2} &=& {\sigma_1}^2\,\f{\Sigma_1}{\Sigma} 
 			\pa{1+\f{\Sigma_2}{\Sigma_1}
 			    \,\f{{\sigma_2}^2}{{\sigma_1}^2}}	\nonumber\\
 			 &=& {\sigma_1}^2\,\f{\Sigma_1}{\Sigma} 
 			\pa{1+\f{\Sigma_2}{\Sigma_1}
 			    \,\f{m_1}{m_2}}.
 \end{eqnarray}
 On the other hand, using light to weigh the quantities yields
 \begin{equation}
	 \sigma\e{lw}^{\ 2} =  {\sigma_1}^2\,\f{\Lambda_1}{\Lambda}
				\pa{1+\f{\Lambda_2}{\Lambda_1}
				    \,\f{m_1}{m_2}}.
 \end{equation}
 Since the stellar IMF is peaked at the low-mass end and $m_1 < m_2$, the
surface density of the second component $\Sigma_2 < \Sigma_1 \approx \Sigma$.
The same quantities weighted by light yield a different result. Since
$m_1 < m_2$, the brightness $\Lambda_1 \ll \Lambda_2 \approx \Lambda$, so that
generally $\Lambda_2\,m_1 > \Lambda_1\,m_2$ for a standard IMF (see 
\S\ref{brightness:comparison}). This gives the following,
approximate, relations
 \begin{equation}
 \moyenne{\sigma^2}  \approx {\sigma_1}^2 \qetq \sigma\e{lw}^{\ 2}  \approx 
	{\sigma_2}^2.
 \end{equation}

 By the same line of arguments, we obtain for the radii
 \begin{equation}
	R\e{hm}	\approx R\e{hm1}	\qetq
	R\e{hl}	\approx R\e{hm2}
 \end{equation}

Owing to mass segregation, we anticipate 
${\sigma_1}^2 > {\sigma_2}^2$ and $R\e{hm1} > R\e{hm2}$. As a result, $\eta$
computed from Eq.~\eqref{eta:eta0} gives
 \begin{equation}
	\eta = \eta_0 \f{R\e{hm1}\,{\sigma_1}^2}{R\e{hm2}\,{\sigma_2}^2} > \eta_0.
 \end{equation}

This reasoning will hold true for multi-mass cases in general since we may
replace $m_1$ by $\langle m\rangle$ and $m_2$ by $m\e{max}$: light-weighted
quantities trace positions and velocities of the heavier stars, whereas the
global potential and kinematics are set by the less-massive stars.

\section{Numerical method}\label{gaseous:model}

\subsection{Gas models}

 Large stellar systems share several thermodynamical properties with classical
gases (Lynden-Bell \& Wood 1968). A cluster composed of stars of different
masses may be likened to a set of concentric spheres of ideal gas satisfying
Poisson's equation.  Larson (1970) pioneered a method based on moments of the
Boltzmann equation by which energy (`heat') flows through the system as it
would in a fluid.  The temperature of a stellar system, then, is identified
with the local square velocity dispersion so that heat may be transported from
low dispersion regions to high dispersion regions (owing to the negative heat
capacity of gravity).  Stellar collisions are treated through a local heat
conduction equation (Lynden-Bell \& Eggleton 1980) which may be calibrated to
give evolutionary tracks virtually indistinguishable from those obtained from
N-body calculations (see Spurzem \& Takahashi 1995).

 Bettwieser
 \& Inagaki (1985) give a good insight of the hydrodynamical spirit of the
model but note that their closure equation requires modification for agreement
with Fokker-Planck models (Spurzem \& Takahashi 1995). A complete and
anisotropic formulation based on moments of the Boltzmann equation can be
found in Louis \& Spurzem (1991).

\begin{figure*}

\plotone{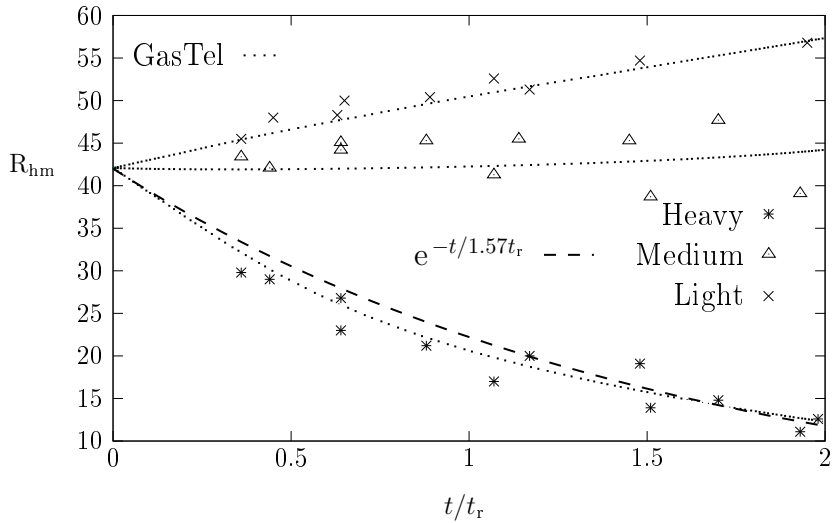}

\caption{Run of half-mass radius for a three-component Plummer model versus
time.  The three stellar masses $2/5$, $1$ and $5/2$ were drawn from a
Salpeter IMF.  The symbols are for data points lifted from Fig.~1 of Spitzer
\& Shull (1975). The dash is an exponential decay $\propto
\exp\pa{-\tf{t}{1.57t\e{r}}}$. \label{Spitzer:Shull}}

\end{figure*}

\subsection{Integration code} 

    The numerical code Spedi\footnote{Further details at 
\texttt{http://www.ari.uni-heidelberg.de/gaseous-model/}} that we use is based 
largely on the formulation for
anisotropic stellar systems due to Louis \& Spurzem (1991). It was developed
further by Spurzem \& Takahashi (1995).
 The equations are set on a logarithmic mesh using a scheme which is
forward-differencing in space and centered in time.  Time-integration was
performed iteratively using a semi-implicit Newton-Raphson-Henyey method. 
     The gravitational potential is evaluated from the updated
(total) density profile directly from Poisson's equation. 

Spedi has been adapted by one of us (Deiters 2001) to include a model of
stellar evolution. We refer to the resulting code as GasTel. Stars are evolved
according to the Cambridge stellar evolution tracks, which are
available in a convenient analytical form (Pols et al. 1998, Hurley et al.
2000). By the end of their lives, stars have lost a significant fraction of
their mass. This mass lost by stars is expelled instantaneously from the
cluster. However, we may still compute $\eta$ in the approximation that the
total cluster mass remains constant for short evolution times since in reality
the gas will not leave the cluster instantly (see \S\ref{long:evolution}).

     All variables are evaluated on a 200-point grid in Heggie \& Mathieu's
(1986) Nbody units. The constant logarithmic width between two grid points is
$\dd \ln{r}\approx 0.095$.
 Spatial resolution in the centre is excellent (152 mesh points to the initial
half-mass radius at $0.6$ numerical units) and the grid extends up to $60$
numerical units.

\subsection{Calibration, tests}

The only free parameter in the equations of the gaseous model is the value of
the conductivity (sometimes denoted $\lambda$). It is then adjusted to be
consistent with N-body calculations and to recover the core collapse time in
the case of a system of $N$ identical masses (Bettwieser \& Inagaki 1985,
\S2.2, see also Spurzem 1992). Several tests have been conducted to compare
gas cluster models with Fokker-Planck integrations and direct N-body
calculations (Giersz \& Heggie 1994; Giersz \& Spurzem 1994 ; Spurzem \&
Takahashi 1995).  Most of these tests were for two-component models of modest
mass ratios, while the stellar mass range of interest here covers nearly two
decades.  We therefore checked explicitly that the numerical setup correctly
reproduces the dynamics of multi-mass models with a broad mass spectrum.
Spitzer \& Shull (1975) presented results of mass segregation from
Fokker-Planck calculations of three-component Plummer models. The three masses
were in the ratio $2/5$:$1$:$5/2$ and drawn from a Salpeter IMF. 
Fig.~\ref{Spitzer:Shull} graphs the time evolution of the half-mass radius of
each component as obtained with GasTel (dotted curves) along with the results
read off Fig.~1 of Spitzer \& Shull (1975).  We find very good agreement with
their data. In particular we find the evolution for the most massive stars
well recovered from an exponential decay of the form $R\e{hm}(t) = R\e{hm}(0)
\exp(-t/1.57 t\e{rh})$.

\section{Method of analysis and error estimates}\label{erreurs}

 GasTel computes for each dynamical mass group the mass distribution and
velocity dispersion on a radial grid.  The mass density $\rho(r)$ and velocity
dispersion $\sigma(r)$ of each group are integrated along the line of sight to
obtain the projected distributions at cylindrical radius $R$ using the
relation \hbox{$r^2 = R^2 + z^2$}:
 \begin{eqnarray}
		\Sigma(R) &=& \Int{-\infty}{+\infty}\rho(r)\:\dd z \\
		\Sigma\,\sigma\e{los}^{\ 2}(R) &=& 
			\Int{-\infty}{+\infty}\rho(r)\,\sigma^2(r)\:\dd z\, .
 \end{eqnarray}
 The half-mass radius and mean velocity dispersion are computed for each mass
group.  System averages are then computed by summing over all groups using
either the density or the light flux as statistical weight. 
 A stand-alone programme is used to pick the most luminous stellar mass at
each output time.  This approach allows us to combine the properties of
different stars as desired in the analysis, without having to re-run the
simulation with a different setup. 
    For example, it will prove illuminating in the first instance to monitor
physical quantities attached to the most luminous stars alone as function of
time, before profiling the system including contributions from all the stars
(see \S\ref{dominant:contribution}). 

\subsection{Mass sampling}
   Our overall goal is to describe accurately the early evolution of a
cluster.  This suggests that we seek out a relation between the spectrum of
stellar masses and the time-scale for dynamical evolution given
by Eq.~\eqref{segregation:time}, before selecting a set of stellar mass 
groups.

\subsubsection{Stellar IMF}\label{IMF}

 The field stellar IMF in the solar neighbourhood sets a standard reference
(Kroupa 2002).  This distribution function of single stars of mass $m$ is well
fitted by a piece-wise power law,
 \begin{equation}
 f(m) \propto \left\{\begin{array}{ll}
			{m}^{-\alpha}	&\text{if }	m 	< 		1\solarm	\\
			{m}^{-\beta}	&\text{if }	1\solarm	< m < 	10\solarm	\\
			{m}^{-\gamma}	&\text{if }	m 	> 		10\solarm	\\
			\end{array}\right.
			\label{eq:IMF} 
 \end{equation}
 where $\alpha=1.30$,
$\beta=2.35$, and $\gamma=4.0$. The
value of $\alpha$ has significant 
uncertainties $\pm0.7$ (Kroupa 2002) whose implications will be discussed in 
\S\ref{sect:IMF:Alpha}.
 Stellar demographics are computed by integrating $f(m)\, \dd m$ up from a
lower value which we set above the brown dwarf limit at $0.10\solarm$.
If the real mass distribution extended to $0\solarm$ with the same power law, 
our cut at $0.1\solarm$ would allow us to account for $80$ per cent of the 
actual 
mass and $99.9$ per cent of the actual emitted light of the low mass stars 
($M<1\solarm$).  The mean stellar mass computed from Eq.~\eqref{eq:IMF}
\label{brightness:comparison} is
 \begin{equation}
	\moyenne{m} = \f{\Int{0.1\solarm}{\infty} m\,f(m)\,\dd m}
	       {\Int{0.1\solarm}{\infty}    f(m)\,\dd m}  \approx  0.7\solarm. 
 \label{mean:mass}
 \end{equation}
 However, note that due to discrete and non uniform sampling, in most of the
simulations we have $\moyenne{m} \approx 0.85\solarm$.

	\subsubsection{The mass spectrum and the importance of stellar evolution}
\label{K:sect}
 The mean value given by Eq.~\eqref{mean:mass} is only weakly dependent on 
the upper bound of integration. 
That upper bound should be chosen so as to reflect the richness of stellar
populations of massive clusters, yet without overwhelming the computational
scheme. 
  
 The lifetime of a star is a steep function of its mass. We find the following 
polynomial to give  a good fit to stellar  lifetimes in the mass range 
$[5\solarm,70\solarm]$: 
 \begin{equation}
	\log t\e{life} = c_1\times\pa{\log(m)}^2 + c_2\times\log m + c_3
	\label{tlife}
 \end{equation}
 where $c_1\approx0.96$, $c_2\approx-3.7$ and $c_3\approx4.2$; $m$ is 
expressed in solar masses, and $t\e{life}$ in Myr.
  A star will take full part in  the time-evolution of the cluster through two-body
scattering if its lifetime exceeds the mass-segregation time 
of Eq.~\eqref{segregation:time}, which we rewrite as (dropping the subscript 
max) 
 \begin{equation}
	t\e{ms} = \f{K}{m}	\label{K:def}
 \end{equation}
 where
 \begin{equation}
	K \equiv \f{\pi}{3}\, \f{\bar{\rho}}{\rho}\, \langle m\rangle
	\pac{\f{r\e{hm}}{r\e{g}}}^{3/2}\,t\e{r}\ . 
 \end{equation}
 Combining the two time-scales allows us to find  a 
 reference mass
 to satisfy the relation 
 $ \log t\e{ms} < \log t\e{life}\ . $ Substituting $t\e{life}$
from Eq.~\eqref{tlife} we obtain
 \begin{equation}
	0  < c_1\times\pa{\log(m)}^2 + (c_2+1)\times\log m + c_3-\log K\:, 
	\label{trinome}
 \end{equation}
 a quadratic inequality for $\log m$. Solving for the roots of this quadratic
we obtain

 \begin{equation}	\log{m^\pm}= \f{-c_2-1}{2\,c_1} \pm 
\f{\sqrt{(c_2+1)^2-4\,c_1\,(c_3-\log K)}}{2\,c_1}\ .  \end{equation}
 The interpretation of this result is straightforward. All stars with initial mass $m \in
[m^-,m^+]$ will not migrate much to the centre
of the cluster in the course of their lifetime on the main sequence. 
Those with masses above ${m}^+$ and below ${m}^-$, will.  Therefore 
to model accurately the very early stages of clusters we should ideally include 
all stars above ${m}^+$. 
  Recently, Figer (2005) has argued from Arches cluster data that all stars
have initially a mass $<150\solarm$: this would set an absolute upper limit on
the mass spectrum. However the impact of such very massive stars on the
dynamics is small since they carry a minute fraction of the total mass and 
luminosity of the
system.
 Thus, the most massive stars we have included in some of the calculations in 
this
paper had $m = 70 \solarm$ which already exceeds the mass of Wolf-Rayet stars. 

 The relation of ${m}^\pm$ to cluster parameters is summed up in the
constant $K$: the fraction of all the stars that will contribute more effectively to 
mass segregation is, therefore, an implicit function
of the cluster we wish to model. The minimum of Eq.~\eqref{trinome} occurs for 
 \begin{equation} 
	\log m = - \f{c_2+1}{2c_1} \simeq 1.41\:, \label{minimum:trinome}
 \end{equation}
 or $m \approx 25.5 \solarm$. This is the only root to the quadratic when 
 \begin{equation} K = c_3 - \f{(c_2+1)^2}{4 c_1} \simeq 200.0 \equiv K\e{c} \end{equation} 
 where the numerical value follows from our choice of fitting parameters but
is otherwise uniquely defined. 

 The meaning of $K\e{c}$ becomes clear if we recall the definition of $K$ and
$t\e{r}$, Eq.~\eqref{K:def} and~\eqref{relaxation:time}. 
 Note first that there are no real roots to Eq.~\eqref{trinome} when $K < 
K\e{c}$. 
Whenever that is the case, all stars drawn from the IMF effectively segregate
while on the main sequence and lose very little mass in the process.  When $K$
exceeds $K\e{c}$, all stars in the interval $[m^-,m^+]$ must evolve significantly on
their way to the centre.  

 But since $K$ increases with the relaxation time $t\e{r}$, 
 itself a rising function of the number of stars $N$ (at a given crossing
time), we may work out a value for $N$ beyond which it is unrealistic to
neglect stellar evolution. 
 We find after some algebra that the condition $K \ge K\e{c}$ reduces to
 \begin{equation}
	\f{\bar{\rho}}{\rho} \f{\langle m\rangle}{\solarm}\, \f{t\e{cr}}{1 {\rm
	Myr}}\,\f{N}{\ln 0.4 N} \ge 86\: K\e{c} \label{N:Kc}
 \end{equation}
 where we have substituted the numerical factor $r\e{hm}/r\e{g} = 0.4$.  Any
multi-mass cluster model (N-body or otherwise) that does satisfy this
inequality and neglects the stellar evolution processes is in error. We can
isolate for $N$ in Eq.~\eqref{N:Kc} by taking characteristic values for the
crossing time and mean stellar mass to be $t\e{cr}\approx 0.5\U{Myr}$ and
$\langle m\rangle = 0.7 \solarm$. If the stars are not segregated by mass
initially then on average $\bar{\rho}/\rho =1$ by definition. With these
values inserted in Eq.~\eqref{N:Kc} we find $N \ge 6 \times 10^5 \equiv 
N\e{c}$,
above the census of an average star cluster in the Galaxy
($\moyenne{M}=300\,000\solarm$), but not atypical of clusters in the
Antenn\ae{} (Mengel et al. 2002, table~3).

 Note that for a given crossing time, a multi-mass calculation with $N<N\e{c}$
will correctly reproduce the migration of stars up to a time of order
$t\e{ms}$ even without accounting for stellar evolution.
 Portegies Zwart \& McMillan (2002) and G\"urkan et al. (2004) used this
argument to model runaway collapse of a cluster leading to the formation of an
intermediate mass black hole. To perform their simulations, they supposed that
the relaxation time of their cluster was less than $30\U{Myr}$ so that
collapse occurs before the first stars explode around $3\U{Myr}$.

\subsubsection{The choice of mass bins} 
\label{lower:mass}

 To include very-low mass stars in the computation is costly and brings little
in terms of the time-variation of the light curves,
cf.~Eq.~\eqref{segregation:time}. The difficulty resides in having to resolve
the light profile of the high-mass stars, and the potential, dominated by
sub-solar mass stars, simultaneously. We mitigated this problem partly by
selecting a low-mass cutoff of $0.1 \solarm$, well below the mean mass
of Eq.~\eqref{mean:mass}.  We then defined the mass ensemble $\{ m_j \}$, $ j 
= 
1,2,
.. J$, such that the
 lifetimes of two successive high-mass components, $m_j, m_{j+1}$, 
  differ by $\approx 5\U{Myr}$ using Eq.~\eqref{tlife}. 
  We refer to all stars with initial mass $m > 5 \solarm$ as `massive
stars'.  In contrast, the sampling of the mass spectrum in the interval
$[0.2, 5] \solarm$ followed a geometric mass-doubling progression:
$m_1=0.2\solarm$, $m_2=0.4\solarm$, $m_3=0.8\solarm$, etc.  We define $m_{1/2}
= m_1/2 = 0.1\solarm$. 
 In short, the ensemble $\{ m_j\}$ spans the mass range $0.2\solarm$ to
$m\e{J}$ non-uniformly and allows a much-improved focus on the evolution of
the massive stars. 

\begin{figure*}

\plotone{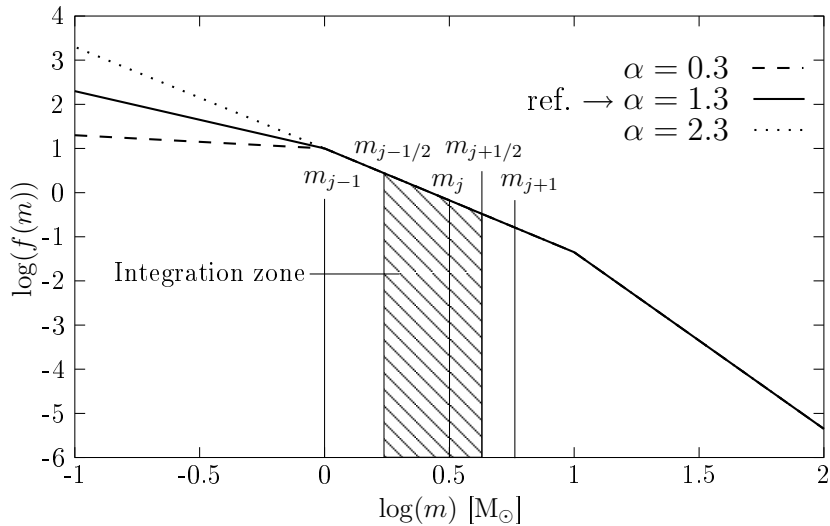}

\caption{Generic form of the stellar IMF, Eq.~\eqref{eq:IMF}. The reference
IMF (solid line) corresponds to the power index triplet $(\alpha,\beta,\gamma)
= (1.30,2.35,4.00)$. Changing $\alpha$ changes the slope on the left side of
the figure. The integration algorithm is illustrated for a mass $m_j$: the
integration boundaries are chosen to be the geometric means $m_{j-1/2} =
\sqrt{m_{j-1}\,m_j}$ and $m_{j+1/2} = \sqrt{m_j\,m_{j+1}}$ which are
mid-logarithmic intervals.  \label{IMF:Alpha}}

\end{figure*}

 With mass bins so chosen, the IMF is integrated over each interval
$m_{j+1/2}-m_{j-1/2}$ to distribute the mass within each bin $j$ and
normalised so that
 
 \begin{equation} \Int{m_{1/2}}{m_{J+1/2}} f(m) \dd m = N\ .  \end{equation} 
The geometric mean of two successive  mass groups has been used to define the
bounds of integration ($m_{j\pm1/2}$) for each group (see 
Fig.~\ref{IMF:Alpha}). 

\begin{figure*}

\plotone{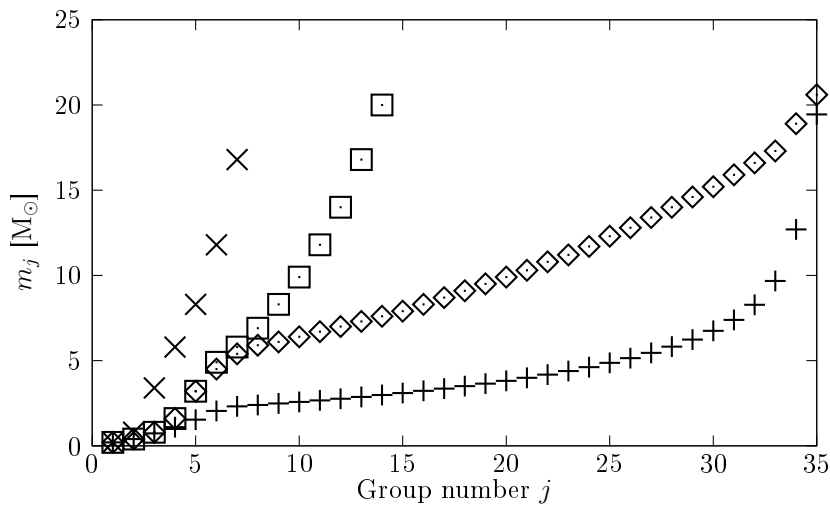}

\caption{Mean value $m_j$ of each mass group $j$ for simulations with 7
($\times$), 14 ($\boxdot$) and 35 ($\DiamondJJ$) components. The
sampling used for the $500\U{Myr}$ evolution simulation (Fig.~\ref{300Myrs})
is also plotted ($+$).\label{fig:mass:groups}}

\end{figure*}

 Simulations were done with 7, 14 and 35 mass groups to investigate variations
due to a finer sampling of the stellar mass function (see
Fig.~\ref{fig:mass:groups}).  The trends in mass segregation are robust to
decreasing or increasing the number of groups, but are more noisy in
calculations performed with on the order of only a few groups. All results in
this article are for runs with $J = 35$ mass groups unless stated otherwise.

\subsection{Radius determinations and Monte-Carlo checks} 

We may distinguish between the half-mass radius derived for the
continuous density profile of the gas model, and the same radius derived for
an N-body rendition of that continuum. 
 
The half-mass radius for each component of the gas model is computed by 
integrating once over
the entire plane to obtain the total mass of the group; the grid is then
re-sampled to identify the radius $R\e{hm}$ enclosing half of the mass. A
linear interpolation at the grid points bracketing $R\e{hm}$ gives accuracy to
second order in the grid interval.  Owing to a very fine meshing up to and
beyond the half-mass radius, errors on this radius
 are
 negligible.  In practice one would like to know what errors are introduced
when a finite-N model is projected on the grid and the same radius evaluated
from star counts.  This is particularly important when the number of stars of
a given mass group is low and statistical fluctuations comparatively large. 

To that end, we performed two sets of Monte-Carlo (MC) tests. 
 First, we computed the half-mass radius for an ensemble of $N$ stars from the
surface density of a Plummer sphere projected on the sky. We call the result
$R\e{mc} $.
 Looking at the dependence in $N$ of the fluctuations of $R\e{mc}$ around
$R\e{hm}$, we concluded that they were of a Poissonian form, mostly due to the
random selection of a star in the given density function. For example, with
$N=50$ stars, the dispersion around the mean value is of order $15$ per cent,
for $N=1000$, it is $3$ per cent.  Second, we perform 1000 Monte-Carlo
realizations of the $500\,000$ stars reference model (see
Table~\ref{initial:conditions}) using the density functions at 10 and
$40\U{Myr}$ to see which are the dispersions of the fluctuations on the mean
value of the half-\emph{light} radius. In both case, the dispersion was of
order $3$ per cent assuring that in a real cluster of half a million stars the
half-\emph{light} radius is dominated by the $1000$ or so (i.e. $0.2$ per 
cent) most luminous stars.

\begin{table*}
\begin{center}
\begin{tabular}{*{8}{c}|*{6}{c}}
\multicolumn{8}{c}{\null\dotfill{} Adjustable quantities \dotfill \null} &
\multicolumn{6}{c}{\null\dotfill{} Derived quantities \dotfill \null} \\
$J$ & $N$ & 
$R\e{hl}\null_0$ & 
$m\e{min}$ & $m\e{max}$  & $\alpha$ & $\beta$ & $\gamma$ &
$M$ & 
$\sigma\e{los}\null_0$ & 
$\eta_0$	&	$\Sigma(0)$	&	$t\e{r}$	&	$t\e{ms}$
\\
	&	&
[pc] & 
[M$\solar$]	& [M$\solar$] & & & &
[M$\solar$] & 
[$\mathrm{km\,s^{-1}}$] &
			&	[$\mathrm{M_\odot\,pc^{-2}}$]	&	[Myr]	&	[Myr]	
\\ \tableline 
$35$& $500\,000$ & 
$1$ & 
$0.2 $ & $20.0$ & 1.30 & 2.35 & 4.00 &
$418\,000$ & 
$15$ & 
8.6 &	$0.67 \times 10^5$ &	$180$ & $2$
\\
\end{tabular}

\end{center}
\caption{Parameters and useful data for the reference Plummer model. $J$ is 
the total number of groups, $N$ the number of stars, $M$ the initial total 
mass. $R\e{hm0}$, $\sigma\e{los0}$ and $\eta_0$ are respectively the initial 
half-mass radius, line-of-sight velocity dispersion and $\eta$. 
$\Sigma(0)$ is the central surface density of the cluster, $t\e{r}$ and 
$t\e{ms}$ its relaxation and segregation times. 
Note that for $J=35$, we have $m_{\tf{1}{2}}=0.1\solarm$ and 
$m_{35+\tf{1}{2}}=20.6$. We also have $m\e{min}\equiv m_1$ and 
$m\e{max}\equiv m_{35}$. The indexes $\alpha$, $\beta$ and $\gamma$ define the 
IMF (see Eq.~[\ref{eq:IMF}]). The 8 parameters on the left are adjustable in 
the code whereas the 6 on the right are derived from them.
\label{initial:conditions}}
\end{table*}

\subsection{Predominant group approximation}
\label{dominant:contribution}

Following
 the latter remark, and in order to have a better understanding of 
what is going on, we decided to restrain our measures to the most luminous 
component whose half-mass radius and velocity dispersion are assumed to be 
the measured half-light radius $R\e{hl}$ and velocity dispersion 
$\sigma\e{los}$ of the whole cluster.

\begin{figure*}

\plotone{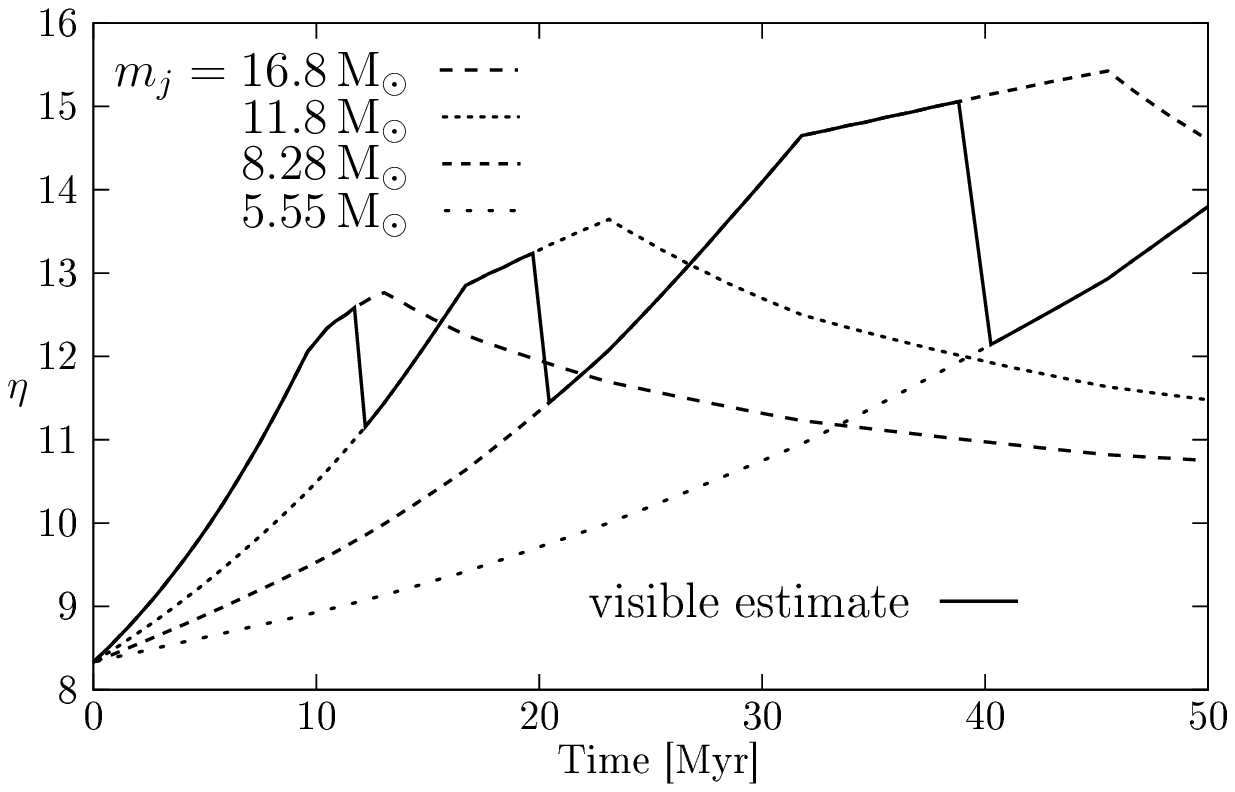}

\caption{Global $\eta$ estimated from the brightest of 7 mass groups. Dashed
and dotted curves show $\eta$ for each group of mass $m_j$ as indicated. The
solid line is the visible estimate for $\eta$ chosen as the brightest
component at any given time. The sudden jumps coincide with the stellar life
time of individual groups.
\label{jumps}}

\end{figure*}

 Figure~\ref{jumps} graphs $\eta$ for each individual group (thin lines) and
the brightest stars (thick solid line). The individual thin curves all
increase from their value at $t = 0$. The increase is steeper for the most
massive stars, as expected.  At $t \approx 10\U{Myr}$, these stars become
supernov\ae{} and turn to faint stellar remnants thereafter.  The $\eta$ we
compute for the system drops sharply to the underlying value given by the new
most-luminous stellar population. And so on for each subsequent episode of
mass loss through supernov\ae{} events. Note that at later times the $\eta$ of
individual stellar groups decreases as a result of significant mass loss. 
This trend is again driven by mass-segregation, when the lighter remnants are
expelled from the central region by the massive stars. Such remnants behave
like point-sources of gravity with no further stellar evolution. The trend
where $\eta$ decreases is caused mainly by an increase of the half-mass
radius, which is more significant than variations in velocity dispersion. 

The value of $\eta$ computed from the most luminous component gives an upper
limit on the value of $\eta$ derived from the integrated light of all stars.
 An illustration based on bolometric light will be discussed in
\S~\ref{long:evolution} (Fig.~\ref{300Myrs}).  Note that the difference
between the two values depends on the wavelength of observation. At the
near-infrared wavelengths often used to study highly reddened starburst
clusters, both values of $\eta$ are essentially identical as soon as the most
massive stars have evolved off the main sequence. Indeed, short-lived red
supergiants or AGB stars overwhelm other sources of light; at each time, their
distribution is that of the currently most massive
objects.\label{eta:lum:less}

\subsection{Number of components}

  The large oscillations seen on Fig.~\ref{jumps} for a model with $J = 7$
components suggested to us to aim for a significantly larger sampling of the
mass function to reduce noise to acceptable levels. 
 Boily et al. (2005) had found for a different reference model that $J=14$
already gave enough precision to identify global trends.  Some uncertainties
remain with $J=14$ models in the later stages of evolution and in particular
when the evolution time exceeds $100\U{Myr}$. 
 This suggested to us to increase $J$ to the largest possible value. 
 The computational time however is quadratic in $J$ and so after some
experiments, we settled for a compromise value of $J = 35$ mass groups binned
inhomogeneously as described in \S4.1. 
 The difference between $J=14$ and $J=35$ models lies mainly in a much
smoother transition at the time when the massive stars undergo rapid mass-loss
while the dynamics for the same mass-component is less affected.

\section{Parameter survey}\label{results}

 We now survey different parameter values for $N$, $m\e{max}$, $R\e{hm0}$ and
$\moyenne{m}$.
 We tried in each comparison to maintain all but one parameter fixed to the
reference model values given in Table~\ref{initial:conditions}. From equations
\eqref{segregation:time}, \eqref{relaxation:time} and~\eqref{crossing:time}
coupled with Eq.~\eqref{virial:theorem} to eliminate $\sigma$, we
get
 \begin{eqnarray}
	t\e{ms} &\propto &
		\f{\moyenne{m}}{m\e{max}}
			\times\f{N}{\ln(0.4\,N)}
			\times\f{R\e{hm0}}{\sqrt{\tf{\moyenne{m}\,N}{R\e{hm0}}}} 
\nonumber\\ 
			&\propto&
		\f{\sqrt{\moyenne{m}\,N\,R\e{hm0}^{\ 3}}}{m\e{max}\,\ln(0.4\,N)}.
		\label{tms}
 \end{eqnarray}
 The survey will highlight dependencies of $t\e{ms}$ on each quantities.
 In all the graphs that follow the solid line indicates the reference model of
Table~\ref{initial:conditions} unless stated otherwise. 
  
\subsection{Particle number}
\label{N}

\begin{figure*}

\plotone{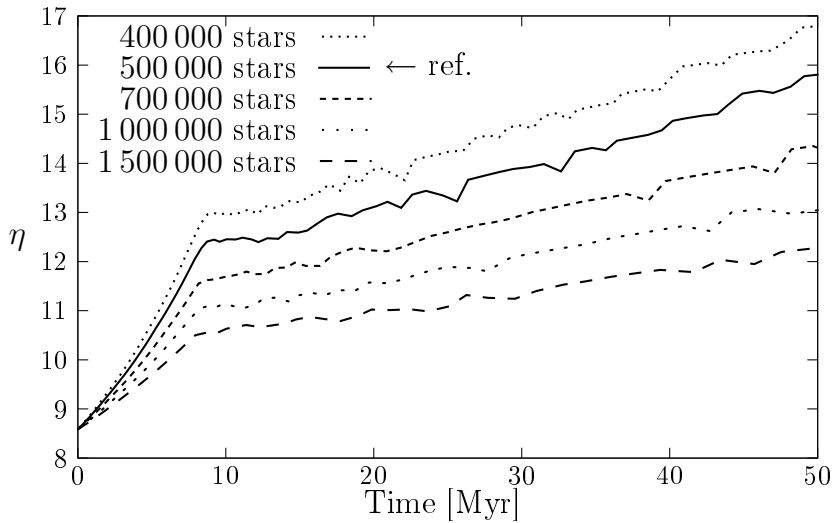}

\caption{Evolution of $\eta$ for different total number of stars $N$.  
 Note the very similar slopes of the curves at times $t > 10\U{Myr}$. The
stellar IMF was truncated at $20\solarm$.  \label{NTOT}}
\end{figure*}
  
  We first investigate the behaviour of $\eta$ when changing the total number
of stars, $N$ while the initial half-mass radius $R\e{hm0}$ is kept unchanged. 
The number $N$ of the survey ranged from 4 to 15 $\times 10^5$ stars. These
values of $N$ bracket the clusters of mass equal to the mean mass of Milky Way
clusters (some $300\,000\solarm$, Meylan \& Heggie 1997) and the very rich
Antenn\ae{} clusters of more than $10^6\solarm$ (e.g. Mengel et al. 2002).  
The
models all have identical mass groups and upper mass limit. With the main
sequence lifetime of $20\solarm$ stars $\approx 10\U{Myr}$ we expect from
Eq.~\eqref{tms}
 a more segregated profile and larger $\eta$ at that time for smaller-$N$
systems, and a similar trend in evolution thereafter.  Fig.~\ref{NTOT}
graphs $\eta$ for five values of $N$ in the range indicated. It is clear from
that figure that richer clusters, with a longer segregation time, show a
less-rapidly changing $\eta$.
 This situation carries over beyond $t \approx 10 \U{Myr}$ when the first
supernov\ae{} events occur, as the second most heavy stars continue to 
converge
to the centre on their own segregation time-scale, also $\propto 
\sqrt{N}$.
 As a result the mass profiles are less segregated when the stars move off the
main sequence in succession for runs with higher values of $N$. Overall
differences in the profiling of $\eta$ at times $t > 10 \U{Myr}$ remains 
small:
for instance the average slope $\dd\eta/\dd t $ is $\approx 1.20$ for the $N =
1.5 \times 10^6$ model, and $\approx 1.31$ for the smallest-N model shown
here. The differences are without major implications if we are concerned with
clusters of ages $< 100 \U{Myr}$ or so. 

\begin{figure*}

\plotone{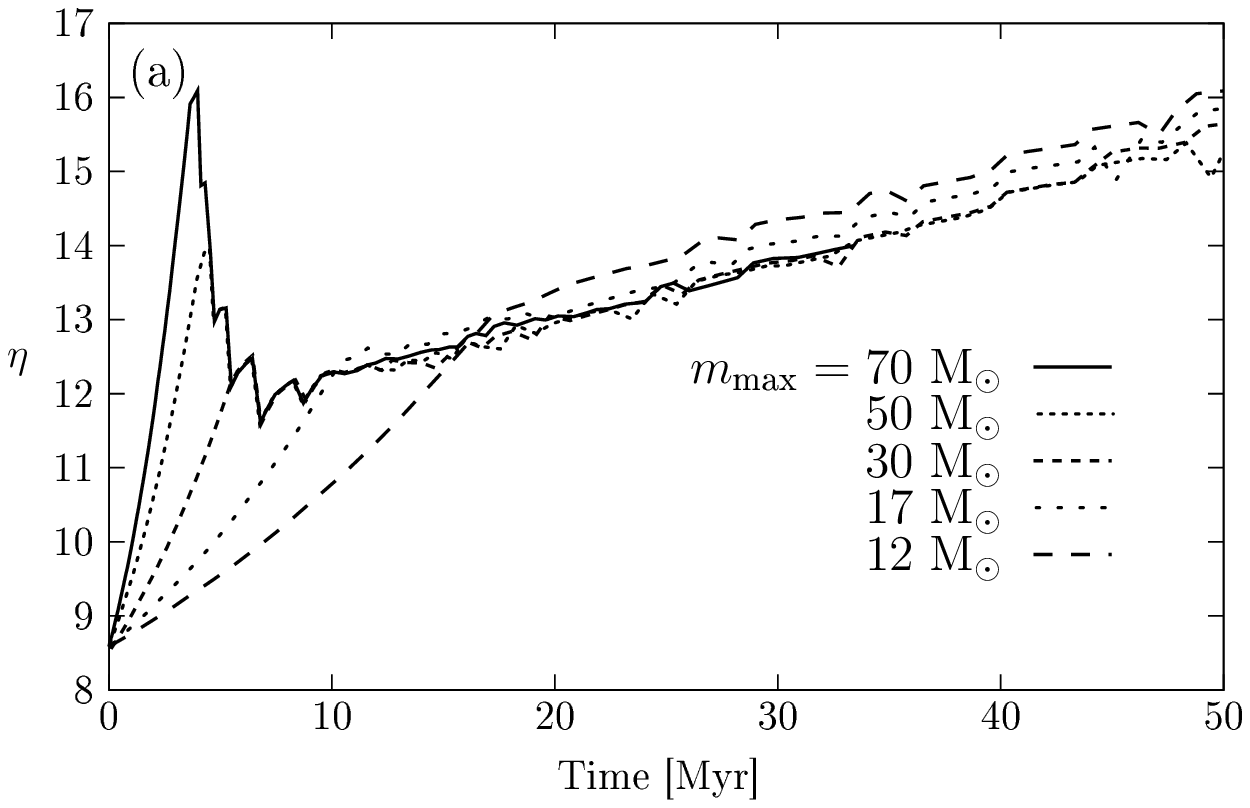}
\plotone{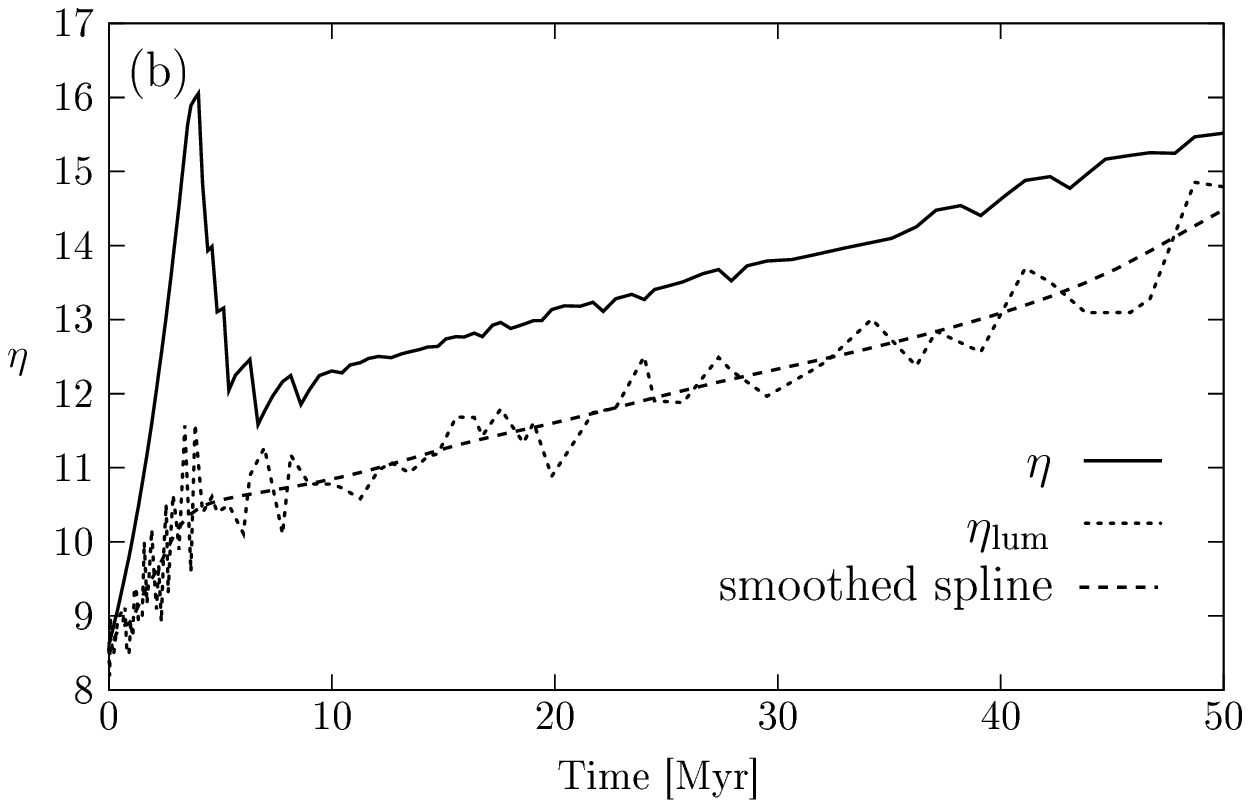}

\caption{Evolution of $\eta$ for models with different upper-mass cutoff
$m\e{max}$.  (a) We set $\eta = \eta_j$ of the brightest component at time $t$
(see \S\ref{dominant:contribution}).  When the stellar mass function is
extended beyond
 $M=25\solarm$, $\eta_j$ is not monotonic and becomes rapidly very large
 at early times. (b) Same as (a) for $m\e{max}=70\solarm$ (solid line) and
$\eta\e{lum}$ computed from summing the light from all the stars at all times
for the same model (dotted line). 
 Note that the first `bump' due to an extended mass range is much attenuated
when computed with total bolometric light curves from MC sampling. The effect
of very bright and short lived stellar states (e.g. AGB) cause many
oscillations due to the random sampling in the profile of $\eta\e{lum}$. 
\label{max:mass}}
	
\end{figure*}

\subsection{Non monotonic evolution ?}

Large-$N$ clusters will host a very rich stellar population and very massive
stars. These stars have very large luminosity but are extremely short-lived;
their impact on the value of $\eta$ should therefore be more significant
on short time-scales.  It is interesting, then, to follow the behaviour of
$\eta$ for individual components when the mass spectrum includes heavy stars
easily identifiable from spectroscopy which could be taken as tracers for the
global dynamics. 
 A tracer might be the brightest stellar component at any given time and we
have seen how $\eta$ can be estimated from the brightest component alone
(Fig.~\ref{jumps}). Would $\eta$ increase monotonically in time if such a
tracer was used instead of a global value obtained from integrated light? 
 Call $\eta_j$ the value of $\eta$ computed for a single component $j =
1,2... J$, of mass $m_j$, and main sequence lifetime ${t\e{life}}_{j}$.  The
mass of the brightest star is a monotonically decreasing function of $t$ and
if $j$ is the brightest mass group at time $t$ then we would say $\eta =
\eta_j$. 

From basic stellar evolution models we have ${t\e{life}}_{j+1} <
{t\e{life}}_{j}$, and so $\eta = \eta_j$ in the time interval
${t\e{life}}_{j+1}< t < {t\e{life}}_{j}$. Thereafter $\eta = \eta_{j-1}$, and
so on.  We noted that $\eta$ is a growing function of $\tf{t}{t\e{ms}}$ while
the stars are on the main sequence.
 Hence to find out whether $\eta$ will increase or decrease as we switch from
$\eta_j \rightarrow \eta_{j-1}$, it is sufficient to check whether
$\tf{t\e{life}}{t\e{ms}}$ is a decreasing or growing function of stellar mass.
In \S4, we fitted the logarithm of $\tf{t\e{life}}{t\e{ms}}$ with a quadratic
function of $\log m$, Eq.~(\ref{minimum:trinome}).  We found a minimum for the
quadratic (i.e. $\log\tf{t\e{life}}{t\e{ms}}$) at $m\approx25\solarm$.
Therefore, if the current brightest stars have a mass that is larger than
$25\solarm$, we should find $\eta_{j-1}({t\e{life}}_{j-1}) < 
\eta_j({t\e{life}}_{j})$.
 On the contrary, if the current brightest mass is $< 25\solarm$, then
$\eta_{j-1}({t\e{life}}_{j-1}) > \eta_{j}({t\e{life}}_{j})$ and hence
$\eta(t)$ measured from tracers would increase with time. 

We graph on Fig.~\ref{max:mass}(a) the curves of $\eta = \eta_j$ for different
cutoffs of the mass function, ranging from 12 to 70 $\solarm$. It is clear
that very large-mass tracers completely bias the value of $\eta$ to large
values, however they can only do so for very short times running up to
$\approx 10\U{Myr}$.  We note that $\eta_j$ is non-monotonic with time for a
cutoff exceeding $25 \solarm$, as expected.  The mass of a star cluster
derived from massive stellar spectroscopic tracers ($>30\solarm$) is off by a
factor that can exceed $\approx 2$ for the reference setup
(Table~\ref{initial:conditions}). Note, however, that at $t \approx 10
\U{Myr}$ and later, all the curves fall back on the same profile, to within
small fluctuations. For comparison, we plot on Fig.~\ref{max:mass}(b) the
solid curve of Fig.~\ref{max:mass}(a) along with the evolution of
$\eta\e{lum}$, the analog of $\eta$ computed from light-weighted integrated
quantities rather than just from the most luminous component. In practice,
$\eta\e{lum}$ is calculated using Monte-Carlo representations of the simulated
clusters.
 Clearly, the bolometric $\eta\e{lum}<\eta$ at all times, as anticipated from
\S\ref{eta:lum:less}.  If a red filter were applied, $\eta\e{lum}$ would be
weighted predominantly by red giant stars, the brightest population, and hence
the gap between the two curves would close up. The time derivative at
$t\gtrsim 10\U{Myr}$ is almost unchanged, so that the trends in time that we
will derive in a forthcoming section applies to either $\eta$.

\subsection{Mean density via $R\e{hm0}$} 
\label{mean:density}

\begin{figure*}

\plotone{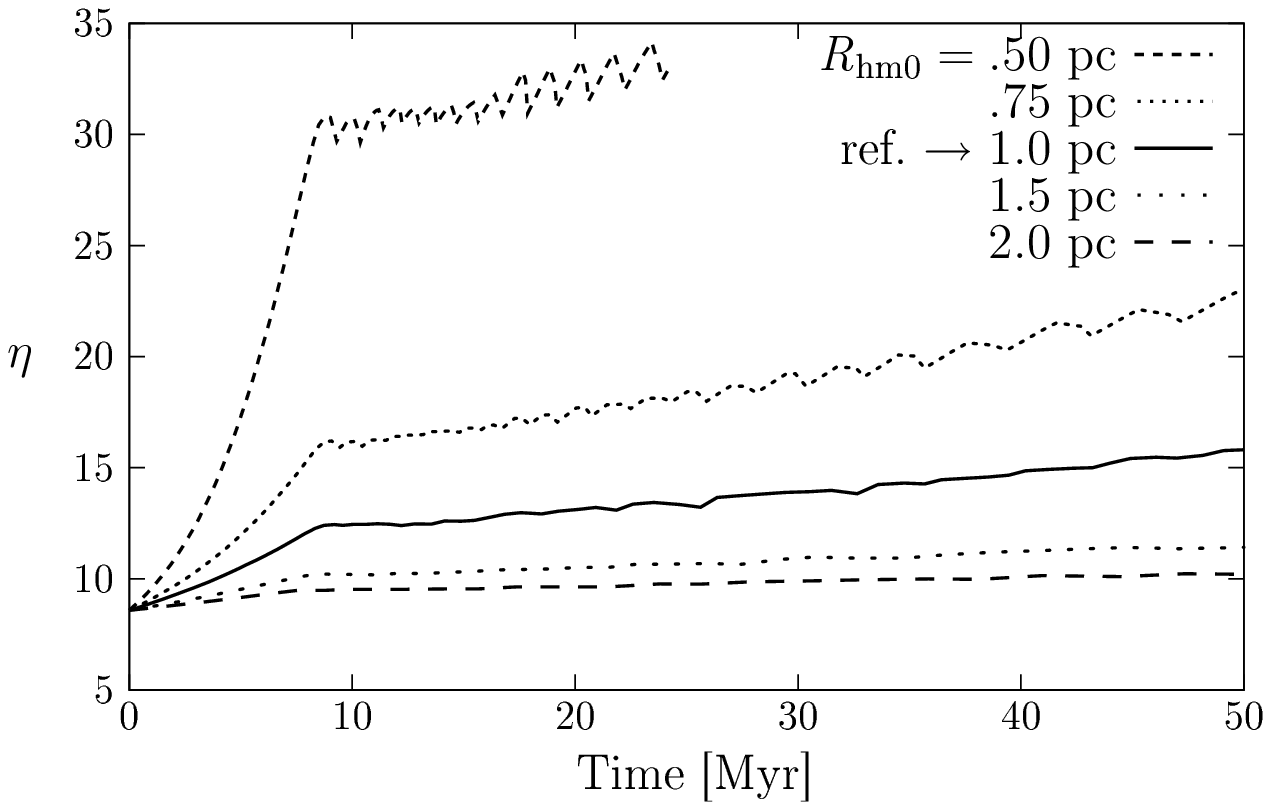}

\caption{Evolution of $\eta$ for models with different projected initial
half-mass radius $R\e{hm0}$.  The denser models have smaller radius. Note that
the low-density model with $R\e{hm0} = 2.0$ pc shows hardly any evolution.
\label{density}}

\end{figure*}

Observed clusters in M82 or the Antennae are compact. They show averaged
surface densities that may exceed the reference value $\approx 7 \times 10^4
\solarm \, {\rm pc}^{-2}$ that we have adopted. Boily et al. (2005) already
noted that the evolution of $\eta$ is significant only for clusters with
surface density exceeding $\sim 10^4 \solarm \, {\rm pc}^{-2}$ (at constant
number of stars). In another context it is known that the mass density of
galactic nuclei may well exceed $10^7 \solarm \, {\rm pc}^{-3}$. 

We have therefore explored the evolution of clusters with different central
surface densities by multiplying lengths by a factor chosen to cover more than
a decade in density.
 The results are plotted on Fig.~\ref{density} for five values of projected
initial half-mass radius, from 0.5 to 2 pc.  The low-density model with
$R\e{hm0} = 2$ pc has a central density $\approx 2\times 10^4 \solarm/{\rm
pc}^2$ and we find an increase in $\eta$ of at most 20 per cent after 
$50\U{Myr}$ of
evolution; by contrast the model with $R\e{hm0} = 1/2$ pc of central density
$\approx 3\times 10^5 \solarm/{\rm pc}^2$ shows a dramatic increase of $\eta$
by a factor $\approx 30/8.6 = 3.5$ in just $10\U{Myr}$ of evolution.  The
rapid increase of $\eta$ is driven by the much shorter dynamical time of the
compact cluster which trickles down to a shorter $t\e{ms}$ in Eq.~\eqref{tms}:
$t\e{cr} \propto R^{3/2}$ implies a segregation time $4^{3/2} = 8$ times
shorter for that model compared with that of the low-density run. 

\subsection{Stellar IMF: $\moyenne{m}$}\label{sect:IMF:Alpha}

There is much on-going debate concerning the universality of the stellar IMF.
The shape of the IMF will fix the mean stellar mass which enters the
definition of the segregation time in Eq.~\eqref{tms}. We already noted 
that stars at the
high-end ($> 10\solarm$) of the mass spectrum carry much light individually
but unless their numbers are greatly enhanced contribute a small fraction of
the total mass. 
 In our exploration of the impact of the shape of the IMF on the dynamics, we
have therefore kept the index $\gamma = 4.0$ as for the reference setup
(Table~\ref{initial:conditions}), and focused instead on the effect of varying 
the low-mass power
index $\alpha$.  As most of the cluster's mass is in low mass stars, $\alpha$
dominates the mean mass value and bears directly on~$\eta$. The same mass
range and number of groups were used in all cases discussed below. 

\begin{figure*}

\plotone{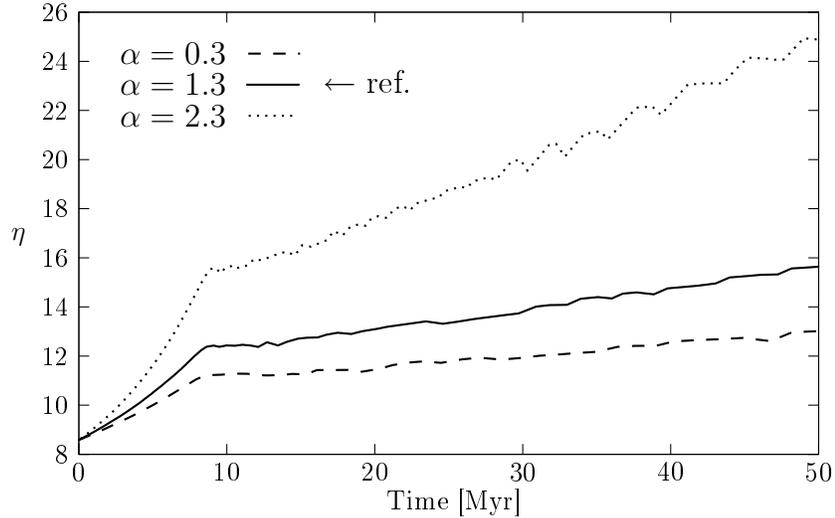}

\caption{Evolution of $\eta$ for the reference model with different low-mass
power index $\alpha$ for the stellar IMF (see Fig.~\ref{IMF:Alpha}). 
The  mean mass are  $\moyenne{m} = 1.3\solarm$,
$0.85\solarm$  and $0.43\solarm$, respectively for 
$\alpha=0.3$, $1.3$ and $2.3$.
\label{influence:Alpha}}

\end{figure*}

Figure~\ref{IMF:Alpha} illustrates the three different IMFs used to perform
the simulations plotted on Fig.~\ref{influence:Alpha}. 
To encompass the standard errors of $\pm0.7$ (Kroupa 2002), we varied our
parameter $\alpha$ by $\pm1$. The chosen upper value $\alpha=2.3 = \beta$
($\moyenne{m} = 0.47\solarm$)  corresponds to a Salpeter profile; it is 
reasonable to assume that any stellar IMF must flatten out at the low mass
end to avoid a divergence in mass. When we do reduce $\alpha$ to 0.3
($\moyenne{m} = 1.3\solarm $), the shift in the evolution of $\eta$ is at
no time as dramatic as the one for the Salpeter value. Flattening the IMF
below the reference $\alpha = 1.3$ profile has not a significant effect on
$\eta$.

\section{Observational implications and long-term evolution}

\subsection{The Stellar Mass Function} \label{sect:mass:function}

 The shape of the stellar mass function might be expected to vary with radius
as a result of mass segregation.
 The central region is rapidly overstaffed with high mass stars while the
outer parts are depleted of them. This trend can be quantified through the
power indices $\alpha$, $\beta$ and $\gamma$ of the mass spectrum,
by comparing the mass function inside and outside a reference radius.
Unfortunately, a cluster evolving rapidly in time offers no fixed
reference radius. To palliate this, we computed the stellar mass function in
two concentric surface elements bounded by the half-light radius from the 
most massive group. While not
specially meaningful, this choice offers the advantage of a direct link
with an observable quantity. 

 From the star counts in each of the 35 mass bins, the mass function is
retrieved by summing all the mass within $m$ and $m+\dd m$, and dividing by
$\dd m$ to obtain a density. We then least-square-fitted power laws in the
ranges $[0.1;1\solarm]$, $[1\solarm;10\solarm]$ and $[10\solarm;20\solarm]$
as in Table~\ref{initial:conditions}. Since the binning is not at constant 
width, we worried
that the mass discretisation would introduce large errors in the values of the
power indices retrieved. To check this, we trained our algorithm on the known
IMF from star counts at $t=0$ for the reference as well as a coarser binning: the
 power indices $\alpha$, $\beta$, $\gamma$ of Table~\ref{initial:conditions} 
were 
recovered to $\pm 0.01$,
which we take as standard deviations. 

\begin{figure*}

\plotone{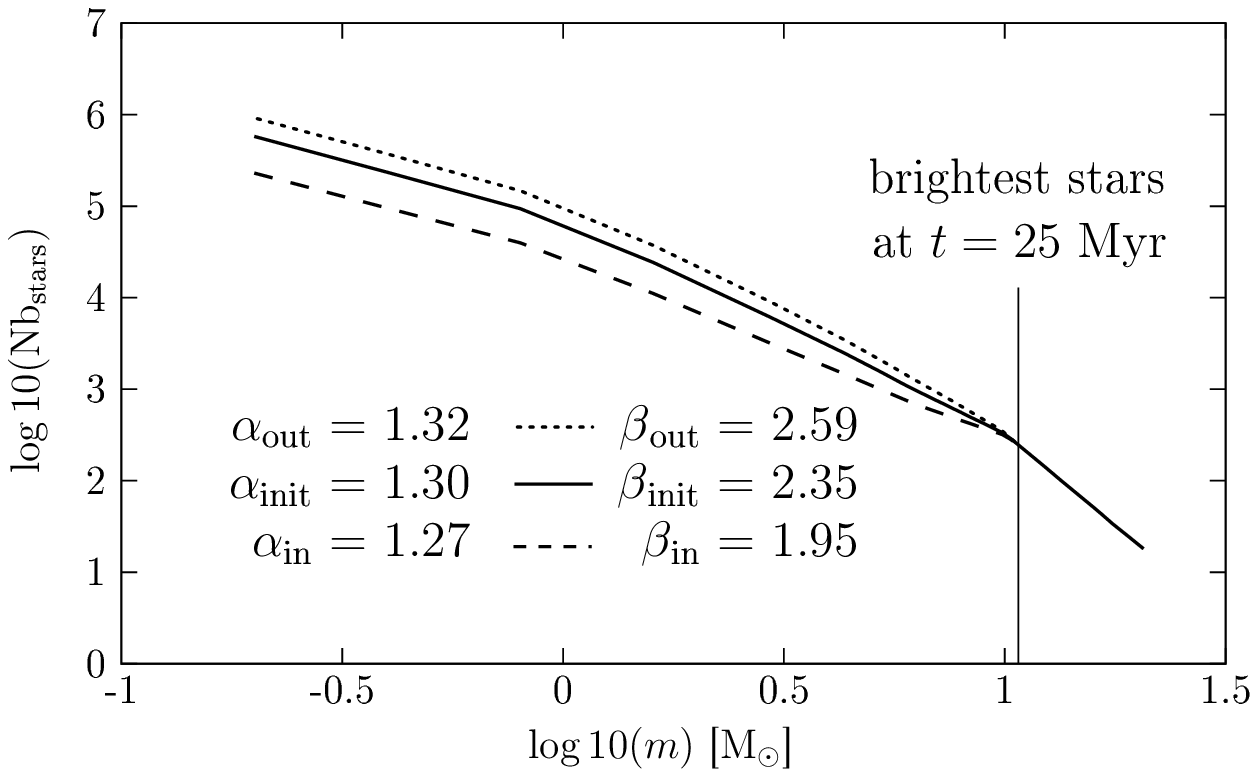}
\plotone{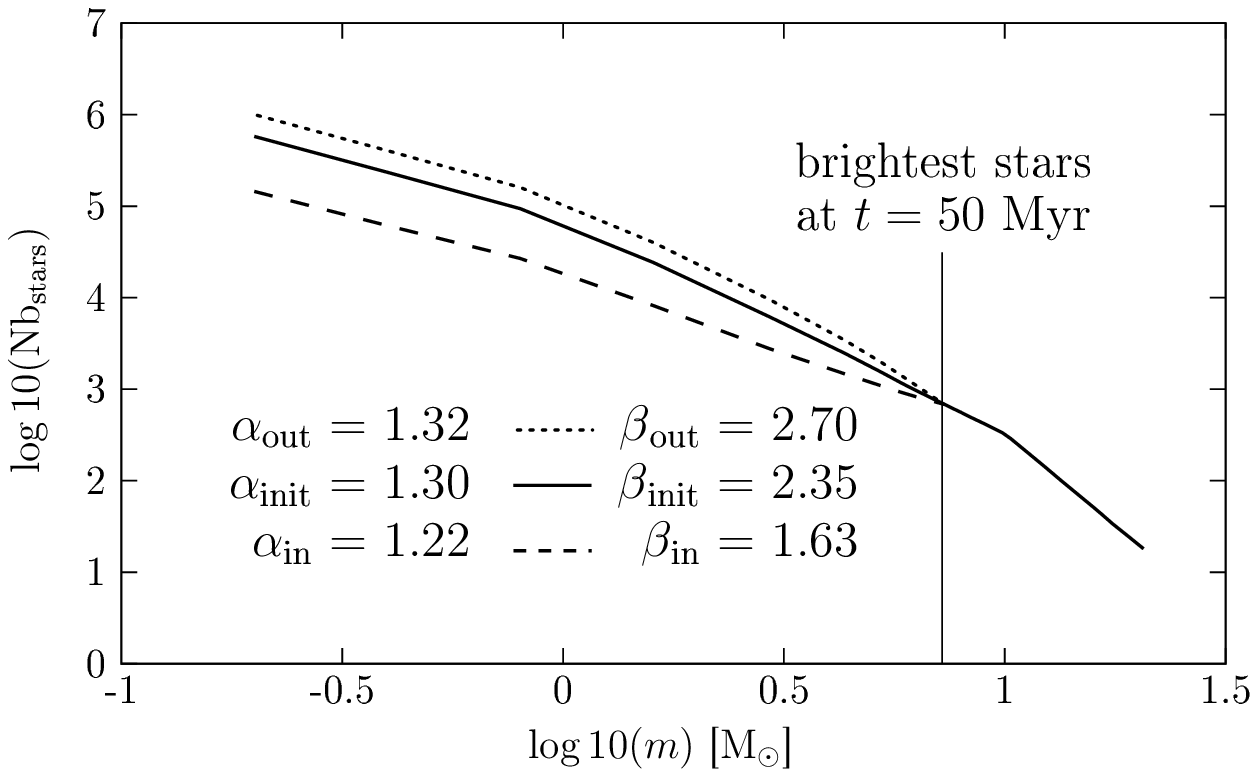}
\plotone{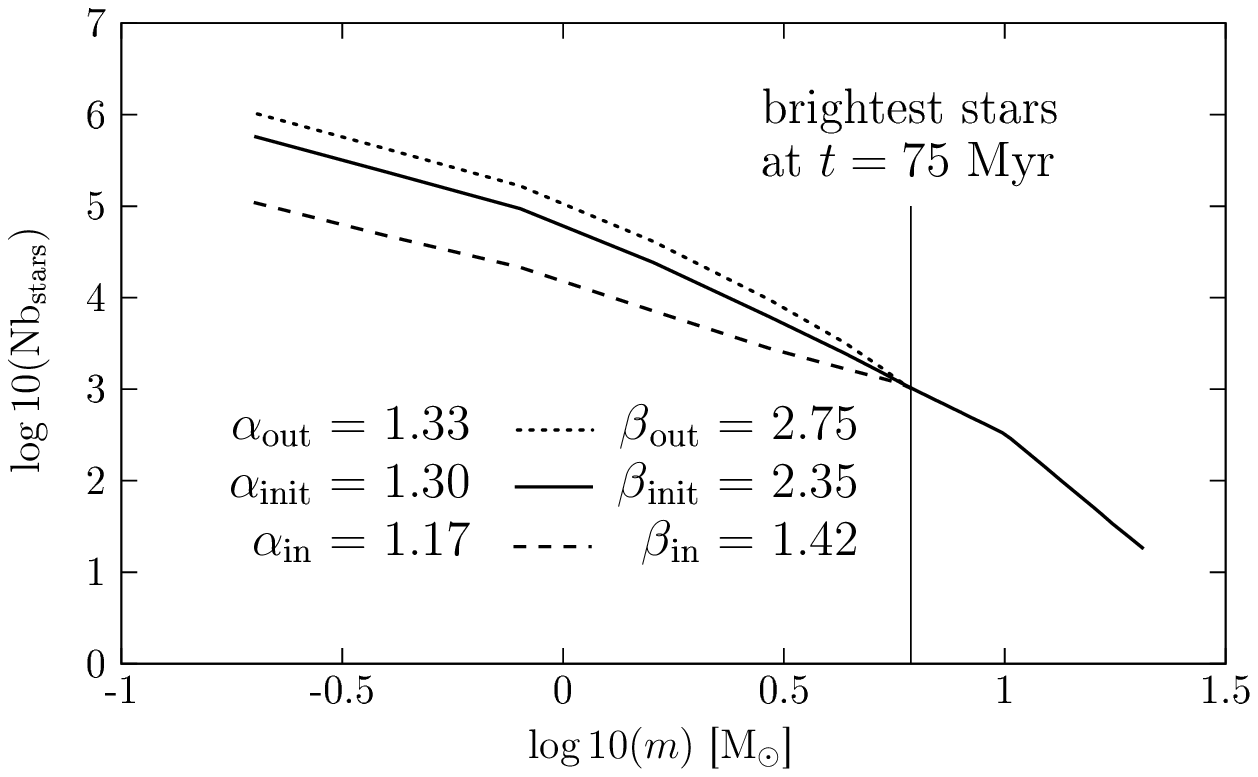}

\caption{The mass function after $25$, $50$ and $75\U{Myr}$ of evolution for a
Plummer model of initial projected half-mass radius $0.75\U{pc}$ (other 
parameters as in Table~\ref{initial:conditions}). The vertical
line indicates the brightest group of stars on each panel. The mass function
has been retrieved, first by computing the projected half-mass radius of the
brightest component, and then by binning stars inside and outside that radius
at each $t$ (see text for details). The slopes $\alpha$ and $\beta$ were
obtained by linear regression of the data on the left of the vertical line.
\label{mass:function}}

\end{figure*}

Fig.~\ref{mass:function} compares the initial mass function (solid line) with
the mass function derived inside (dashed line) and outside (dotted line) the
half-light radius at three different times. With the standard model, the
changes in the mass function are small, therefore the model cluster in this
section was initially twice as concentrated ($R\e{hm0} = 0.75\U{pc}$) as the
reference Plummer model ($R\e{hm0} = 1\U{pc}$) but otherwise the same. 
 The three curves are trivially identical at $t=0$ with slopes given by the
Kroupa IMF. 
 As time increases, the low-mass power index $\alpha$ remains virtually
unchanged in the outer region (varying from 1.3 to $\simeq 1.33$) but shows a
noticeable decrease in the inner part of the system, down by $\approx -0.25$
after $75\U{Myr}$ of evolution. The power index $\beta$ for the mass range
$[1\solarm;10\solarm]$ shows the strongest variations of all, down from its
initial value by as much as $-1$ inside the half-light radius, and up by
$+0.4$ outside this radius. The steeper slope in the outer region is a direct
consequence of the outward migration of light stars initially inside the
half-light radius, while heavy stars flow in the opposite sense. 

For very massive stars, the situation is made slightly more complicated by the
fact that the life-time of these stars is comparable to, or less than, the
evolution times displayed on Fig.~\ref{mass:function}. The vertical straight
line on each panel indicates the mass for which the life-time equals the time
displayed.  All stars to the right of this line are low-mass remnants from
e.g. supernov\ae{} events: these stars therefore do not contribute to the
light profile of the cluster.  Stars initially in this mass range are now
contributing a small addition to the census at the low-mass end of the
distribution. The high-mass part of the diagram is therefore completely
depleted, and has not been fitted. 

These trends with radius are similar to those measured in young LMC clusters
such as NGC 1818 (Hunter et al. 1997; de Grijs et al. 2002b, Gouliermis et al.
2004). This cluster has an age of $\sim 30\U{Myr}$ falling 
between the times displayed on Fig.~\ref{mass:function} and a calculated
relaxation time $\sim 250\U{Myr}$ assuming a half-mass radius of $2.6\U{pc}$ 
and mass of $30\,000\solarm$ (de Grijs et al. 2002a) with 
$\moyenne{m}=0.85\solarm$.
 This is longer than in our simulations.  Note however that the mean density
of the model with $R\e{hm} = 0.75\U{pc}$ matches well the density inside the
half-mass radius of the cluster NGC 1818 (Elson et al. 1987). Inspection of 
Fig.~9
of Gouliermis et al. (2004) shows that the power indices in the inner part
(e.g. $< 0'.3$) derived from their data are similar to those of our
simulation.
This raises the possibility that {\it dynamical } mass segregation
may yet play a key role at the heart of that cluster while
\textit{primordial} segregation is needed to explain the external parts,  
a conclusion already reached by de Grijs et al. (2002b).

\subsection{Colours} 

 It is interesting to investigate to a fuller extent observable consequences
of mass segregation. To that end, we extracted colours from our model clusters
by coupling the Cambridge evolution tracks to the spectral library of Lejeune
et al. (1997, 1998). Nebular gas emission lines, thought present in embedded
young clusters (Anders \& Fritze-v. Alvensleben 2003), are left out of the 
current analysis. 

\begin{figure*}

\plotone{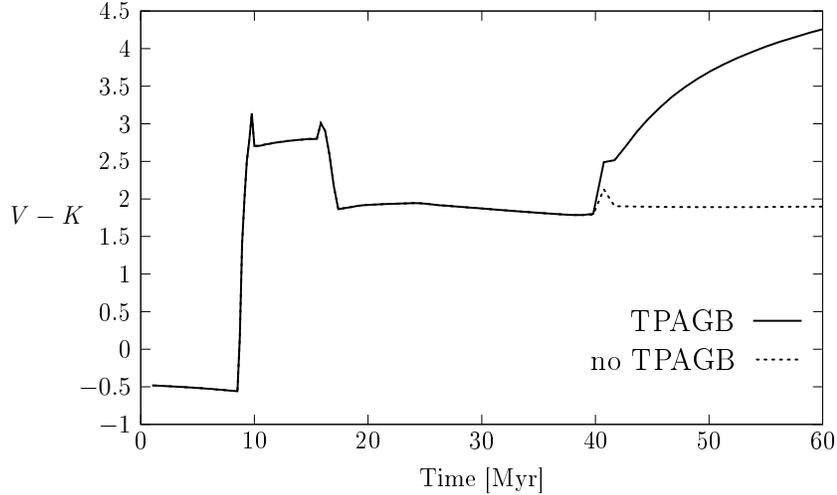}

\caption{$V-K$ plotted against time for the reference model cluster 
(see Table~\ref{initial:conditions}).
  The solid line is for integration with all possible stellar phases
including the TPAGB. The dotted line is the same but without the TPAGB phase.
\label{Couleur:Charlot}}

\end{figure*}

Sampling the mass spectrum
 requires some care in order
to minimize errors in colour magnitudes as discussed by Charlot \& Bruzual
(1991).  First, we tabulated the various evolutionary epochs for a large set
of masses split in equal logarithmic intervals from $2\solarm$ to
$100\solarm$.  The luminosity function was constructed by carefully
integrating the light flux from all the stellar masses in a given evolution
phase, paying great attention to resolve such brief but very bright phases as
the upper asymptotic giant branch (AGB). 

Colours were computed in different wavebands (B, V, I from Bessell 1990,
and K from Bessell \& Brett 1988) and
 compared with those of other authors, who used the evolutionary tracks of the
Geneva group, the Padova group or variations thereof (Girardi \& Bertelli
1998, Bruzual \& Charlot 2003, Mouhcine \& Lan\c{c}on 2003 and references
therein; mostly based on tracks by Bressan et al. 1993 or Schaller et al.
1992). We found that the Cambridge tracks produce significantly redder colours
than others, which have in general been more specifically tuned to reproduce
the observed integrated colours of star clusters.  The origin of these
differences lies in the time stars spend in the late, red phases of stellar
evolution.  In calculations using the Cambridge tracks, the predominance of
red supergiants and upper AGB stars at near-IR wavelengths is probably
exaggerated. To illustrate the effect of the luminous red stars, we
ran simulations with and without stars on the thermally pulsing AGB (TP-AGB).
Fig.~\ref{Couleur:Charlot} shows the evolution of the integrated $V-K$\ colour
of the model cluster. The two curves bracket values commonly found in the
literature for simple stellar populations.

\begin{figure*}

\plotone{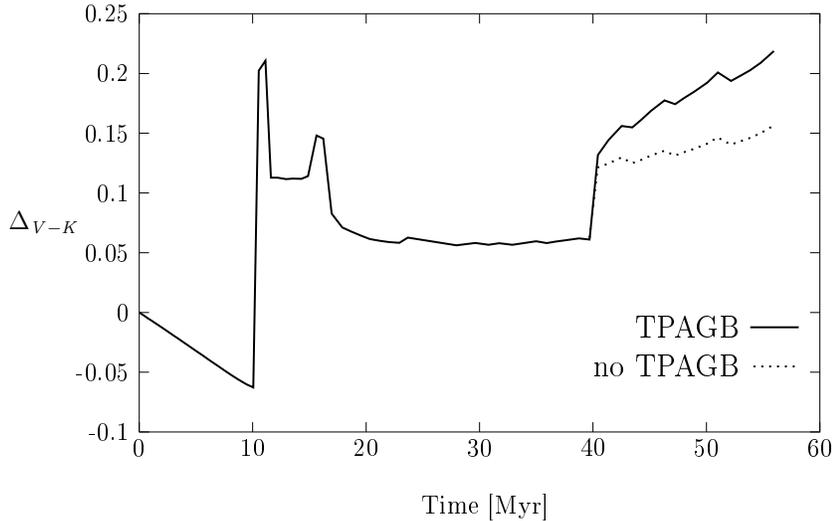}

\caption{Difference of colour $\Delta_{V-K}$ between internal and external
part of a cluster relative to its half-light radius at each given time.  The
inner region becomes bluer over the first $10\U{Myr}$ of evolution, when the
most massive stars in the sample ($m\e{max} = 20\solarm$) become red giants.
 Thereafter the inner part remains systematically redder than the outer region
by as much as 0.2 dex over the first $50\U{Myr}$. Note however that the color
index $\Delta_{V-K}$ fluctuates wildly.
 The trend of increasing $\Delta_{V-K}$ at $t>40\U{Myr}$ is continued in time
exceeding $60\U{Myr}$ (off scale).
 To account for possible bias from the TPAGB phase, when stars are very bright
in the red, we recomputed the colors by removing all stars in that phase of
evolution (dotted line on the figure).
 The behavior stays unchanged. \label{color:grad}}

\end{figure*}

Colour gradients were quantified by measuring the difference between the
colours measured inside and outside the projected  half-light radius:
 	\begin{equation} \Delta_{V-K} \equiv  (V-K)_{< R\e{hl}} - (V-K)_{>R\e{hl}}. \end{equation}
 $\Delta_{V-K}$ is positive when the inner half of the cluster is redder than
the outer half. Fig.~\ref{color:grad} shows the evolution of $\Delta_{V-K}$
for the reference model (Table~\ref{initial:conditions}). At early times, mass
segregation tends to make the centre bluer, as the massive stars that fall
towards the centre are still on the main sequence. Once these stars evolve off
the main sequence, the centre of the cluster very rapidly becomes the reddest
part. Further evolution of the colour gradient presents fluctuations that
reflect the lifetimes of stars of progressively smaller initial masses and the
mass-dependent time they spend as luminous red objects. However,
$\Delta_{V-K}$ remains above 0.05 magnitude throughout, even when the
brightest stars such as those of the TP-AGB are artificially switched off.

\subsection{Long-term evolution}\label{long:evolution}

 It is natural to ask whether the continued increase of $\eta$ observed for
the reference model over a time-scale of $\sim 100\U{Myr}$ carries over to
longer time-scales.  Recall that the relaxation time of that model
 $ \sim 200\U{Myr}$ from Eq.~\eqref{relaxation:time}. We would anticipate some
increase in $\eta$ for any cluster with an age $\gtrsim t\e{r}$, however 
longer
relaxation times also imply longer mass segregation times and significant
contributions from low-mass stars to $\eta$. 
 Since the stellar winds of lower-mass stars moving off the main sequence are
less energetic, it is not clear whether or not the residual gas will be
evacuated on a very short time-scale and how this will impact on the dynamics. 
We take the view that much of the gas can either remain within the cluster for
a period exceeding $t\e{r}$, or on the contrary be evacuated rapidly, thus
 hopefully bracketing all realistic cases. Below we assess whether either
limit (or both) will result in a drop in $\eta$ over long time-scales. 
 
 We ran the reference calculation (Table~\ref{initial:conditions}) for up to
$500\U{Myr}$.  That stretch of time would correspond to several revolutions
through the galactic potential of a galaxy such as the Milky Way, and overrun
the star-burst phase of a typical galaxy merger where young clusters are
observed to form.  Since heavy stars play only a minor role beyond $\sim
30\U{Myr}$ of evolution (cf.~Fig.~\ref{max:mass}), we split the 35~mass bins
so that the stellar MS lifetimes now differ by $\approx 20\U{Myr}$ from one
bin to the next (instead of 5 adopted earlier, see
Fig.~\ref{fig:mass:groups}). In this way the part of the mass spectrum below
$5 \solarm$ is far better sampled than previously.  Nevertheless the
discreteness of the mass function will cause large fluctuations, particularly
noticeable at times $t \gtrsim 200\U{Myr}$ (cf.~\S\ref{dominant:contribution})
and tend to underestimate the increase of $\eta$ as we have seen when
comparing simulations with fewer components. 

\begin{figure*}

\plotone{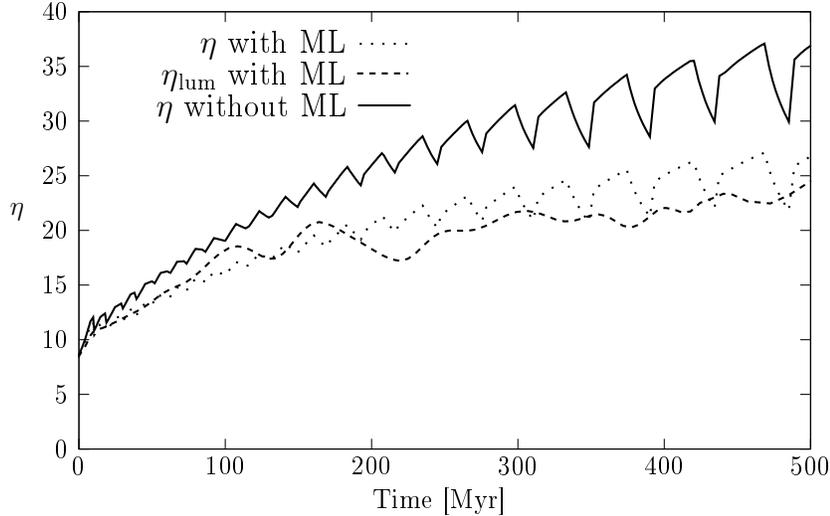}

\caption{Parameter $\eta$ vs time for an 35-component simulation (see text for
details of the numerical setup). The top-most solid curve was obtained by
computing $\eta$ at constant total cluster mass, so disregarding mass loss
(ML) due to stellar winds. The dotted curve assumes on the contrary that this
ML leaves the cluster instantaneously. A more realistic situation where the
stellar winds escape over a finite time interval would pitch $\eta$ somewhere
between the two curves shown here. Note that $\eta$ was once more calculated
by taking the most luminous component as tracer. If we draw Monte Carlo
realisations using the luminosity from all the stars to compute $\eta =
\eta\e{lum}$ under the same assumption of instantaneous ML, we instead obtain
the dashed curve which shadows closely the dotted line. Either way of
computing $\eta$ yields continued increase up to $500\U{Myr}$.\label{300Myrs}}

\end{figure*}

  On Figure~\ref{300Myrs} we graph $\eta$ evaluated in three different ways
for comparison. The dotted curve (denoted `$\eta$ with ML') assumes
instantaneous evacuation of the mass released through stellar evolution. The
dashed curve denoted $\eta\e{lum}$
 also assumes instantaneous mass evacuation but is the analog of $\eta$
computed (as in Fig.\ref{max:mass}) from light-weighted quantities, using a
Monte-Carlo representation of the cluster.
 To verify the impact of stellar mass loss, we also plot the value of
$\eta$ obtained with a constant total cluster mass.
 The result is shown on Figure~\ref{300Myrs} labelled as `$\eta$~without~ML'
(solid line).  For that curve we find an increase of $\eta$ of a factor
ranging from 5 to 6 by the end of the run. This curve remains significantly
and systematically higher than the other two throughout the run. These curves 
bound,
therefore, all possible values because in practice some of the gas will leak
out and $ \eta \propto M $ would decrease as a result. A full simulation
including hydrodynamical effects would trace $\eta$ somewhere between the
curve shown here for $\eta\e{lum}$ and the solid line of $\eta$ without ML. 

It is highly likely that tidal fields will affect the morphology of a cluster
as it orbits the host galaxy. This will certainly change the profiling of
$\eta$ in time, for example by stripping some of the cluster's mass. But
unless the cluster evolves in a very strong tidal field, the dynamics inside
the half mass radius should prove relatively robust. If the total cluster mass
decreases, stripping mass outside $r\e{hl}$, then $\eta$ would decrease from
the values obtained here. 
  A full inspection of this issue would require three-dimensional model
clusters which lay beyond the scope of the current study. 

\section{Application to Cluster Mass Functions}

\subsection{$\eta-t\e{r0}$ relation}\label{sect:prediction}

The results of~\S\ref{N} and~\ref{mean:density} suggest a common thread
linking models of high mean density and those of smallish total mass. Since
these two quantities combine to give the system relaxation
time in Eq.~\eqref{relaxation:time}, we may hope to relate $\eta$ for models 
of
different relaxation times but comparable ages for a given IMF through a
simple scaling formula involving the initial relaxation time $t\e{r0}$.
 Thus we take the view that evolution will be driven almost exclusively by
two-body relaxation and not e.g.  variations in the stellar mass function.
Such effects would play a significant role, as shown on Fig.~\ref{max:mass}, 
but only for
clusters not older than a few million years. 

\begin{figure*}

\plotone{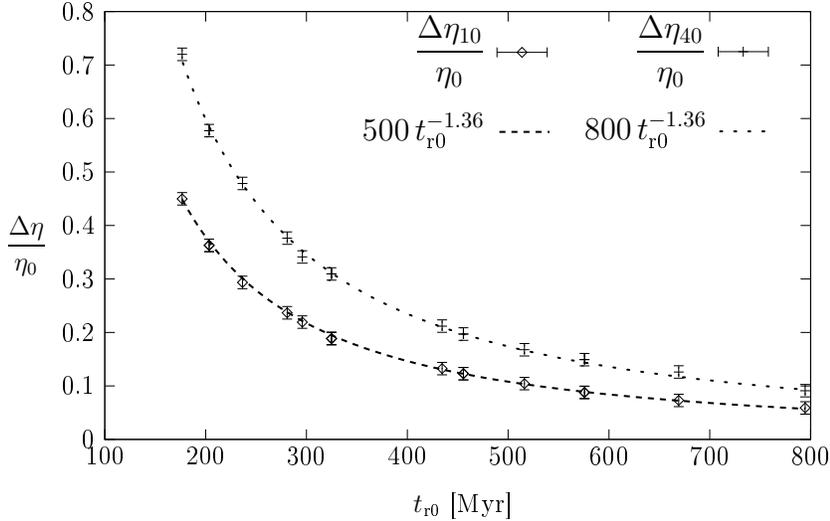}

\caption{The parameter $\eta$ as a function of initial relaxation time
$t\e{r0}$ from different simulations where both $N$ and $R\e{hm0}$ have been
varied.  The relative increase $\Delta\eta/\eta_0$ is a power law of
$t\e{r0}$. The error bars are deviations from the mean value of $\eta$ at each
time. \label{fig:power:laws}}

\end{figure*}

We sought out such a scaling relation by performing a series of runs for
models with different values of $t\e{r0}$, spanning a wide
range of values in $R\e{hm0}$ and $N$ but in other respects identical to the
reference model (Table~\ref{initial:conditions}). 
 We computed $\eta$ for these models at $t=10\U{Myr}$, approximately when the
first stars become supernov\ae{}, and a time $t=40\U{Myr}$ which gives an
intermediate age between e.g. age estimates of Antenn\ae{} clusters (Mengel et
al. 2002) and that of M82-F (Smith \& Gallagher 2001, McCrady et al. 2005). 
Fig.~\ref{fig:power:laws} plots the variations in $\eta$ at these two times
relative to its initial value, $\Delta\eta_t/\eta_0$, as function of
$t\e{r0}$. Both sets of points are well fitted by single power laws,
 \begin{equation}
	\f{\Delta\eta_t}{\eta_0} = A_t\times t\e{r0}^{-a_t} \label{calcul:eta}
 \end{equation}
 and we list the values of $A_t$ and power index $a_t$ in
Table~\ref{power:laws} for two different times. 
 The error bars shown on the figure result from oscillations of $\eta$ 
due to the mass sampling.
 The power-index $a_t$ 
shows only a mild dependence on the age of the cluster.  This is somewhat
surprising if we note that the same power-law functional fit applied to the
half-light radius $\Delta R\e{hl}/R\e{hl0}$ and square velocity dispersion
$\Delta\sigma\e{1d}^{\ 2}/\sigma\e{1d0}^{\ 2}$ give equally good results but
now the parameters vary much between the two chosen times (cf.
Table~\ref{power:laws}). 

\begin{table*}

\begin{center}
\begin{tabular}{c*{2}{|ccc}|}
	&	\multicolumn{3}{c|}{$t=10\U{Myr}$}
	&	\multicolumn{3}{c|}{$t=40\U{Myr}$} \\ 
	&	$\eta$	&	$R\e{hl}$	&	$\sigma\e{1d}^{\ 2}$
	&	$\eta$	&	$R\e{hl}$	&	$\sigma\e{1d}^{\ 2}$	\\
\tableline
$A_t$	& $500$	& $-90$	& $-57$	& $800$	& $-76$	& $-36$	\\
 $a_t$& $1.36$& $1.15$	& $1.22$	& $1.36$& $1.05$	& $1.08$	\\

\end{tabular}
\end{center}

\caption{Power law fits to $\eta$ as function of the initial relaxation time. 
A least-square fit of the functional form defined by Eq.~\eqref{calcul:eta}
was performed at two cluster ages. The standard deviations on $A_t$ and $a_t$
are respectively $5$ per cent and $1$ per cent. \label{power:laws}}

\end{table*}

\begin{figure*}

\begin{center}
\plotone{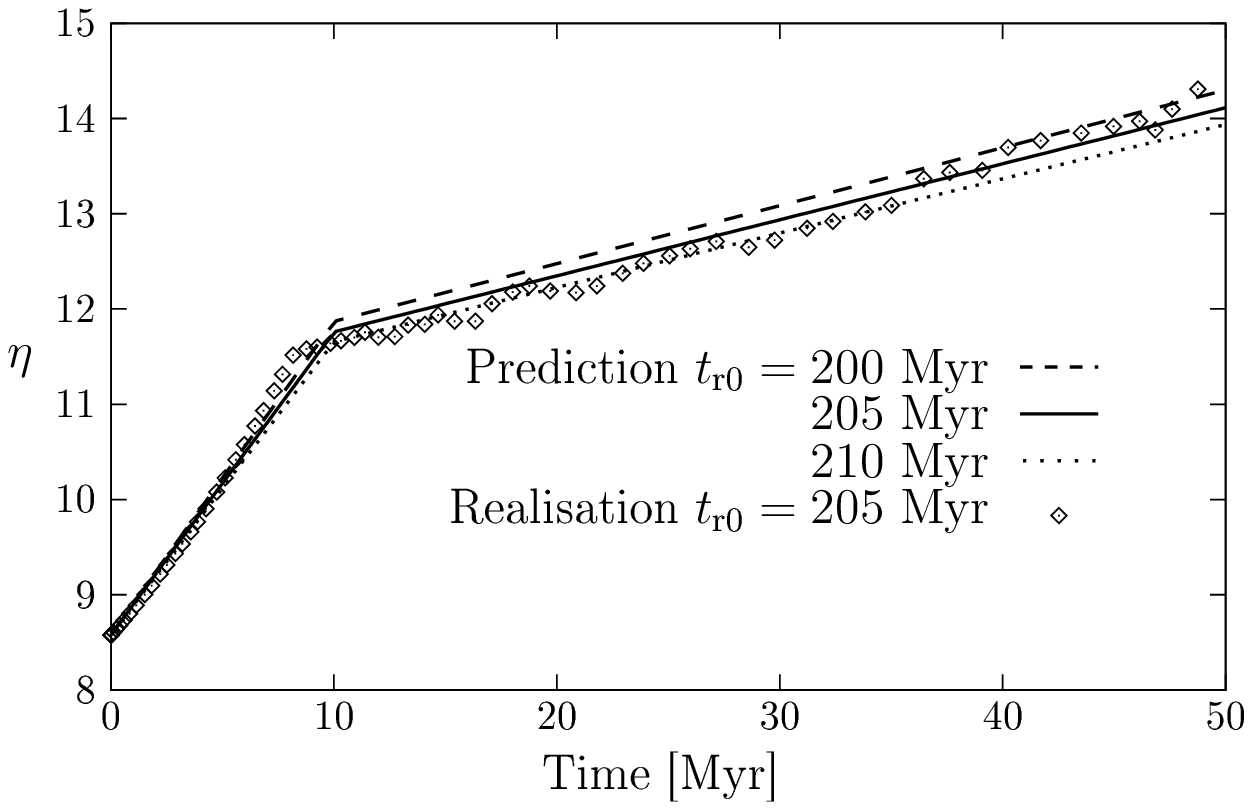}
\end{center}

\caption{Example of the reconstruction of the evolution of $\eta$ in time. The
symbols ($\DiamondJJ$) represent the data from the model while the various
curves are drawn from the parameter fits of Table~\ref{power:laws} using 3
values of $t\e{r0}$ in Eq.~\eqref{calcul:eta}. \label{prediction}}

\end{figure*}

A general expression for $\eta$ valid at ages between 0 and about $50\U{Myr}$
can be derived from the above. Guided by the aspect of Figs.~\ref{NTOT},
\ref{density} or~\ref{influence:Alpha}, we distinguish an early regime of
rapid evolution, up to $\sim 10\U{Myr}$, and a subsequent regime of slower
changes. A good fit to model values is obtained by running a straight line
from the value of $\eta$ at $t=0$ to its value at $t=10\U{Myr}$ for the early
regime, and another straight line through the values at 10 and $40\U{Myr}$ for
the later evolution, using Table~\ref{power:laws}. It can be summarised, with 
$t$ in Myr,
 \begin{equation}
 \begin{cases}
 a_t &= 1.36 \\[2mm]
 A_t &= 	\begin{cases} 
		50\, t 			& 	\text{ if } t < 10\U{Myr} \\
		10\, t + 400	&	\text{ if } t > 10\U{Myr}.
		\end{cases}
 \end{cases}
 \end{equation}
 Fig.~\ref{prediction} compares the analytical values with the results of a
simulation with $t_{\rm r0} \simeq 205\U{Myr}$. Differences do not exceed
10 per cent even when extrapolating to ages of $\sim 100\U{Myr}$. This level 
of
error is found for all models with an initial relaxation time above
$100\U{Myr}$. When the relaxation time is shorter, we find that the
interpolation scheme systematically underestimates $\eta$.

\subsection{Application to a model CMF}

 Using the linear interpolation scheme of the preceding paragraph we may
compute the current mass conversion factor $\eta$ of a cluster from
Eq.~\eqref{calcul:eta} given $t\e{r0}$ and its age $t < 100\U{Myr}$. The real
mass of the cluster so retrieved may then be compared to the mass estimate we
would have computed had we kept the initial value $\eta_0\approx 8.6$ constant
throughout; the ratio of real to estimated mass 
equals $\eta(t\e{r0},t)/\eta_0$. To
compute $\eta(t\e{r0},t)$ first requires a cluster mass and half-mass radius, 
in order
to evaluate $t\e{r0}$ from Eq.~\eqref{relaxation:time}. To do this for an 
ensemble
of clusters, we set up Gaussian distributions for $M$ and $R\e{hm}$ as well as
the cluster age, $t$. These distributions were each sampled independently with
$10^4$ realisations using a standard Ulam-von Neumann (Monte Carlo) rejection
method. 
 In the following, we quote the dispersion $\sigma$ of these distributions as
uncertainties on the mean value (i.e., mean$\, \pm\sigma$). 

\begin{figure*}

\plotone{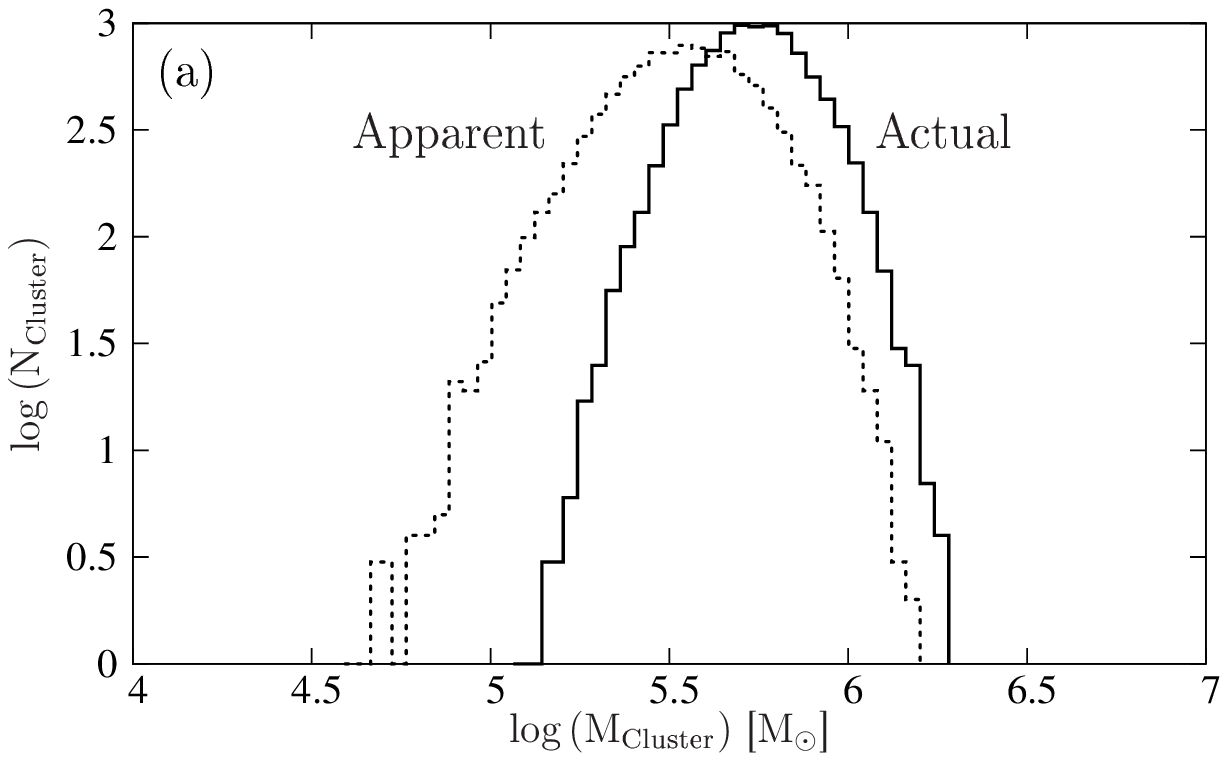} 
\plotone{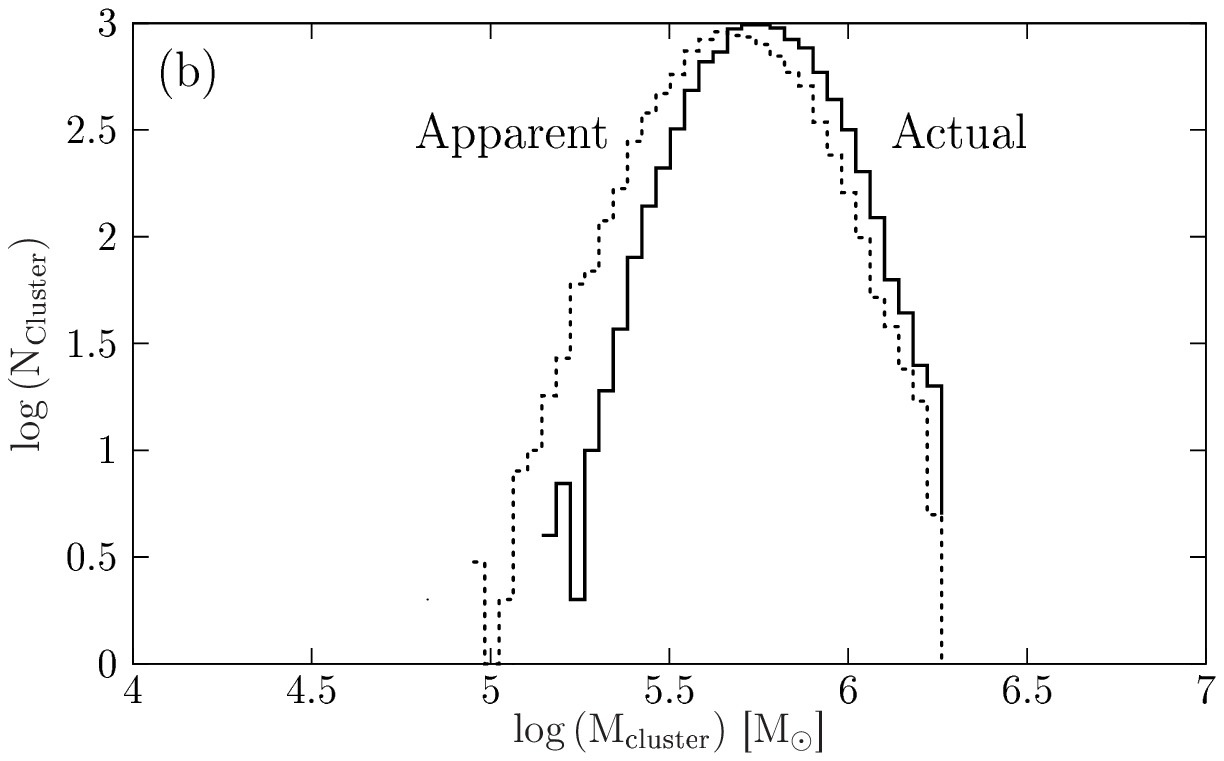}

\caption{Apparent (dotted curves) and actual (solid curves) cluster mass
functions drawn from Gaussian distributions in radii and ages. (a) Results for
a cluster mass distribution of mean radius equals $1\pm0.2\U{pc}$ and mean age
equals $30\pm5\U{Myr}$; and (b) same as (a) but with mean radius equals
$2\pm0.4\U{pc}$ and mean age equals $60\pm10\U{Myr}$. The MC realisations 
contained $10\,000$ clusters.
 The resulting distribution of initial relaxation time are (a)
$190\pm70\U{Myr}$ and (b) $550\pm130\U{Myr}$.  \label{CMF}}

\end{figure*}

 The results are shown on Fig.~\ref{CMF} with a log-Gaussian CMF of mean
mass $5 \times 10^5 \solarm$ for two realisations: (a) an ensemble of compact
clusters of mean radius $1\pm0.2\U{pc}$ and age $30\pm5\U{Myr}$, in line with
values adopted for our reference model; and (b) an ensemble of mean radius
$2\pm0.4\U{pc}$ and mean age $60\pm10\U{Myr}$ inspired from M82-F cluster
data (McCrady et al. 2005).  The resulting relaxation time distributions are
respectively $190\pm70\U{Myr}$ and $550\pm130\U{Myr}$. 

Because the mass derived assuming a constant $\eta = \eta_0$ is always lower
than the actual cluster mass, the distribution shifts
to lower masses as compared to the true CMF. 
 Table~\ref{shift} list the displacements of the peak of the CMF for four
Gaussian distributions of different mean initial relaxation times $t\e{r0}$,
each of the same $50\U{Myr}$ dispersion. It is clear that the shorter 
relaxation
times lead to a larger shift and we find a maximum shift of $0.2\U{dex}$ for
the distribution of average $t\e{r0} = 150\U{Myr}$.
 Very similar conclusions apply for the ensemble of longer-relaxation time 
when the average age is also longer (case [b] above, displayed on the
right-hand panel on Fig.~\ref{CMF}). 

In this spirit the very massive and young Antenn\ae{} clusters are
particularly interesting.  Table~3 of Mengel et al. (2002) lists parameters
for five young clusters of ages ranging from 6 to $10\U{Myr}$ and masses from
$600\,000$ to $5\times10^6\solarm$. The projected half-mass radius of these
clusters is significantly larger than $1\U{pc}$, the value we have adopted for
our reference model. Those large radii may mislead one to expect much larger
relaxation times and, consequently, little or no evolution of the $\eta$ over
time.
 However, we note that Mengel et al. (2002) fitted King models with a
concentration parameter $\Psi/\sigma \approx 6$ to the light profiles of their
clusters. Such King models are significantly more concentrated than the
Plummer model that we used in the calculations performed here. 
 In fact, it turns out that a King model with $\Psi/\sigma =6$, a half-mass
radius of 2 to $3\U{pc}$ (cf Table~3 of Mengel et al. 2002) and a mass of some
$500\,000\solarm$, has a mean density within its core radius that essentially
equals the mean density within the half-mass radius of our reference Plummer
model.
 Consequently, the rapid evolution of~$\eta$ found in this article should be
applicable to the dynamics in the core of some Antenn\ae{} clusters, implying
that the core regions of very rich clusters are still affected by strong
segregation despite their very low age, a conclusion also reached by de Grijs
et al. (2005) using observational arguments.
 As a result, the core will appear
more compact, while the half-mass radius is left largely unchanged, a
situation that leads to King model fits with higher values of $\Psi/\sigma$
than is appropriate (cf. also Boily et al. 2005 for examples of this effect). 

\begin{table*}
\begin{center}
\begin{tabular}{c|cccc}
$\moyenne{t\e{r0}} \pm 50\U{Myr}$	& 150	& 250	& 350	& 450	\\	
\tableline
shift of $\moyenne{\log M}$	& 0.2	& 0.15	& 0.1	& 0.05 \\
\end{tabular}
\end{center}

\caption{Shift of a lognormal distribution in mass centered on $10^6\solarm$
for different mean relaxation times.  The cluster population has a Gaussian
age distribution of mean $30 \pm 20\U{Myr}$.\label{shift}}

\end{table*}

\section{Discussion}

This paper investigated possible biases when estimating the dynamical mass of
young and dense stellar systems from spectro-photometric data.  Using the
virial theorem, one may convert observed half-\textit{light} radius and
flux-weighted mean velocity dispersion to mass through the dimensionless
parameter $\eta$ defined in Eq.~\eqref{virial:observation}. This factor will
vary with time due to mass segregation whenever the two-body relaxation
time in Eq.~\eqref{relaxation:time} is short: 
the heavy bright stars segregate to the centre rapidly.
A parameter space exploration led us to
conclude that for clusters where $t\e{r0} \lesssim 200\U{Myr}$ $\eta$ may
increase by a factor of 2 compared with its initial value, whereas if $t\e{r0}
\gtrsim 500\U{Myr}$ then very little evolution of $\eta$ will take place
within the first $100\U{Myr}$.
 Meanwhile, the mass distribution and potential, dominated by fainter stars, 
remains largely unchanged, so that light does not follow mass anymore.

 We can synthesise the main features of this bias in $\eta$ in a diagram
of cluster age versus relaxation time derived from observations.
 Substituting the projected half-mass radius $R\e{hm}$ and the total mass $M$ 
for 
$r\e{g}$ and $N$ in Eq.~\eqref{relaxation:time}, we get
 \begin{equation}
	t\e{r0} = 
			\f{2\times0.138\,\eta_0^{1/2}}
			  {G\,\moyenne{m}\,\sqrt{3}\,\ln\Lambda}
			\, R\e{hm}^{\ 2}\, \sigma\e{m1d} \label{real:trh}
 \end{equation}
 where $\moyenne{m}$ is given by the IMF and $\ln\Lambda$ is the Coulomb
logarithm (for which we adopted $\ln[0.4\,N]$ previously). The same formula
applied to \emph{observed} quantities would give an estimated relaxation time
 \begin{equation}
	t\e{r,obs} =
			\f{2\times0.138\,\eta_0^{1/2}}
			  {G\,\moyenne{m}\,\sqrt{3}\,\ln\Lambda}
			\, R\e{hl}^{\ 2}\, \sigma\e{los}. \label{observed:trh}
 \end{equation}
 With the ratio $\eta/\eta_0$ given by Eq.~\eqref{eta:eta0} 
 we obtain a ratio of `true' to measured relaxation times, 
 \begin{equation}
	\f{t\e{r0}}{t\e{r,obs}} = \paf{\eta}{\eta_0}^2 
			 \paf{\sigma\e{los}}{\sigma\e{m1d}}^3 \ .  
					\label{relax:ratio}
 \end{equation}
 The ratio of squared velocity dispersion was found not to decrease by more
than 10 per cent throughout the simulation time; we may therefore set
$\sigma\e{los}/\sigma\e{m1d} = 1$ in Eq.~\eqref{relax:ratio}.
 Equation~\eqref{relax:ratio} together with Eq.~\eqref{calcul:eta} can be 
solved
with input cluster age and measured relaxation time to obtain a unique pair
$(t\e{r0},\eta)$. It is then straightforward to draw lines of constant $\eta$
in a graph of age vs $t\e{r,obs}$ axes and recover the true relaxation time of
the mass profile (since we must have $t\e{r0} = t\e{r,obs}$ at time $t = 0$ by
construction and the potential does not change).
 
\begin{figure*}

\plotone{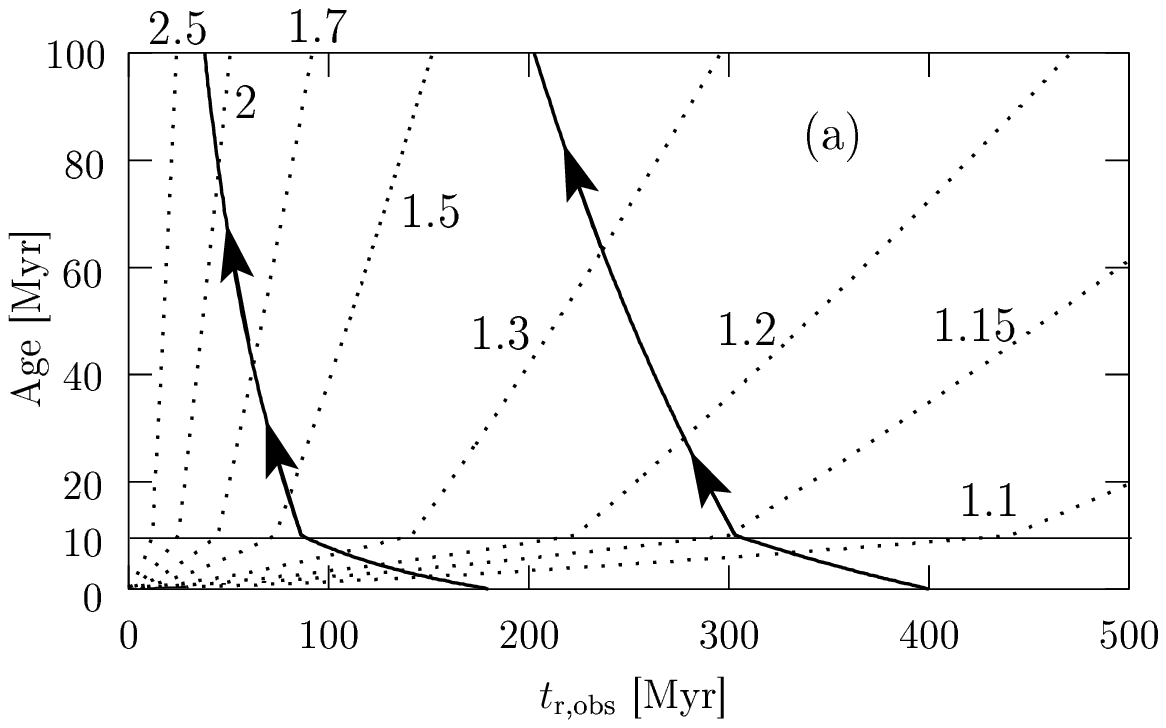}
\plotone{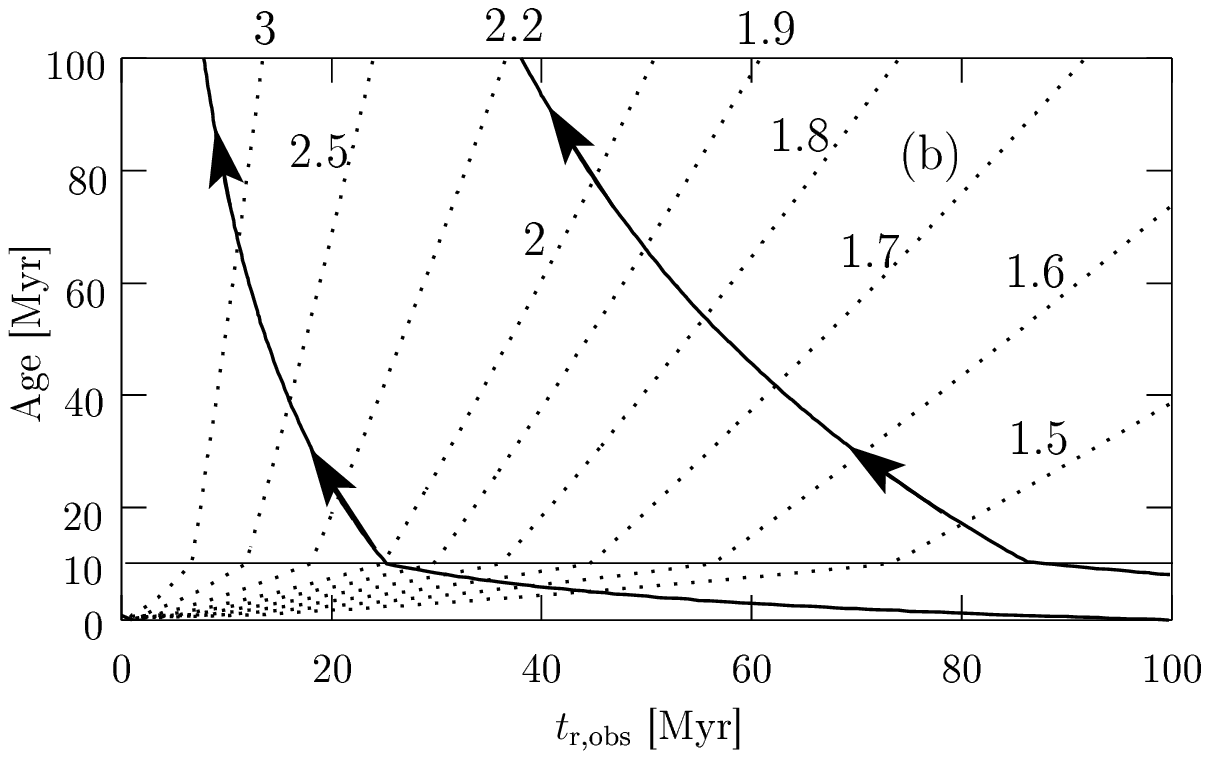}

\caption{Contours of constant $\eta(t)/\eta_0$ (dotted curves) in the plane of
cluster age vs relaxation time, $t\e{r,obs}$, derived from cluster
observables.  The contours were constructed using the bi-linear interpolation
scheme of \S\ref{sect:prediction}. The contour $\eta(0)/\eta_0 = 1$ coincides
with the horizontal axis (cluster age$\null= 0$).  (a) The solid lines trace
the evolution in time of $\eta/\eta_0$ for two model clusters with initially
$t\e{r,obs} = 180$ and $400\U{Myr}$. The arrows point to the future. The value
of $t\e{r,obs}$ decreases with age, always, while $\eta(t)/ \eta_0$ increases.
(b) Same as (a) but now one of the models has $t\e{r,obs} = 100$ Myr
initially. The evolution track for that case crosses contours of yet higher
values of $\eta(t)/\eta_0$ at fixed age. Note the change of scale on the
absciss. \label{eta:diagram}}

\end{figure*}

Fig.~\ref{eta:diagram} graphs the contours lines of constant $\eta/\eta_0$
in the plane $t\e{r,obs}$--age. 
 All clusters start off on the age $ = 0$ axis which coincides with the
contour $\tf{\eta}{\eta_0}=1$.
 As the cluster becomes older and mass segregation sets in, the time
evolution marks a path that is seen to drift to shorter (measured)
relaxation times and larger $\eta$.
 Each level is indicated on the graph. 
 The path for our reference model is shown along with a second model of
initial half-mass relaxation time of $400\U{Myr}$.  Note that even for
this model $t\e{r,obs}$ drops to $\approx 200\U{Myr}$ after $100\U{Myr}$
of evolution: at that time its mass is underestimated by $\approx 30$ per 
cent.  
Fig.~\ref{eta:diagram}(b)  zooms in on the interval $[0:100]\U{Myr}$ of
the $t\e{r,obs}$-axis. 

As seen on Fig.~\ref{eta:diagram}, the most dramatic evolution in $\eta$
occurs in the first $10\U{Myr}$. Furthermore, if only the most luminous
stars were used as tracers, still higher factors $\eta/\eta_0$ would be
expected (cf.~Fig.~\ref{max:mass}).  For the young massive Antenn\ae{}
clusters for which the measured relaxation time $t\e{r,obs} >
500\U{Myr}$, mass segregation is negligible when applied to these
clusters as a whole.  We noted, however, that our reference model provides
a good fit to the core region of some of these clusters in terms of mass
and density. Thus mass segregation may yet prove an efficient agent for
evolution in the central part of massive young clusters, a process that
would contribute to make the core look more compact than it really is and
so inflate the concentration parameter $c = \log (r\e{h}/ r\e{c} )$ of King 
model
fits to these clusters.  Examples of this phenomenon are given in Boily et
al. (2005).

Light curves have been used to estimate colour indices averaged over two
surface elements bounded by the projected half-light radius.  The difference
between these colour indices taken as a function of time shows that the inner
part becomes bluer by 0.05 dex so long as massive stars have not reached the
red giant state. At that time and for all times thereafter, the inner region
becomes redder (by more than 0.05 dex). 
 Colours should be interpreted with caution. 
 The evolution of $V-K$ is highly sensitive to the properties of the red
stages of stellar evolution, and especially to the red supergiant and
asymptotic giant branches. Such fine details of stellar evolution, often
model-dependent and difficult to pin down with precision, have less bearing on
the colour \emph{gradients} because these giants stars dominate the light in
the red wherever they are, and hence $\Delta_{V-K}$ quantifies their
concentration in space. The stellar evolution tracks used here tend to
exaggerate the role
 of bright red
 stars. Discarding the most luminous red supergiants or TP-AGB stars 
altogether reduces the gradients by a few hundredths of a magnitude at most.

 The evolution of the stellar mass function within the model clusters was also
quantified through variations in the power-indices defined for the IMF
Eq.~\eqref{eq:IMF}. We noted that the variations of the power index $\beta$
during evolution are a good match to those observed in the LMC cluster NGC
1818 (Gouliermis et al. 2004). However the initial relaxation time of the more
concentrated model used in this section of $\sim 115\U{Myr}$ is significantly
smaller than the one derived for this cluster (we compute $\sim 250\U{Myr}$
from Eq.~\eqref{relaxation:time}; see also Elson et al. 1987) which implies
that dynamical mass segregation alone does not account for gradients in the
stellar population for the cluster as a whole. Thus we would argue that a fair
degree of primordial stellar segregation must be relied on to explain that
cluster's photometry.  Despite this caveat, it is well worth repeating that
the {\it central} relaxation time of NGC 1818 of some $120\U{Myr}$ is of order
of the half-mass relaxation time of the model used, and therefore dynamical
segregation surely has played a role in the evolution of the stellar mass
function near the centre (see also de Grijs et al. 2002b). 
 Our view is that a set of models tailored to that particular cluster will be
required to disentangle fully primordial from evolutionary effects.

 With $\eta(t\e{r0},t)$ derived from Eq.~\eqref{calcul:eta}, it was possible
to construct a Log-Gaussian cluster mass function and carry out a statistical
survey of the impact of mass segregation on the shape of the observed CMF
obtained from assuming no evolution of $\eta$, to the CMF derived from taking
into account the time-evolution of $\eta$ (Fig.~\ref{eta:diagram}).  The
actual total integrated mass of the CMF is $50$ per cent larger than the 
apparent one
for the case describe on Fig~\ref{CMF}(a) and $20$ per cent for 
Fig~\ref{CMF}(b).
This has direct bearing on the global star formation rate (SFR) derived for
galaxy mergers and starburst galaxies in general. We noted that the two CMF's
so constructed differ mostly at the low-mass range ($<500\,000\solarm$) where
the relaxation time is significantly shorter. To quantify this effect, we
found a shift in the peak of the real CMF towards high masses compared to the
`observed' CMF. This shift is on the order of $0.2$ dex for a distribution
of relaxation times centered around $200\U{Myr}$ and is lower when this mean 
relaxation time is larger (Table~\ref{shift}). Furthermore we
also found a slight widening of the observed CMF, by a logarithmic factor
$\simeq 0.05$ (see Fig.~\ref{CMF}).  This would have some influence on
evolutionary predictions of cluster mass functions.  Vesperini (1998) has
shown that the Milky Way CMF may well shift to {\it smaller} masses by $\simeq
0.1 $ dex over a Hubble time due to tidal destruction and other effects. The
trend we found here goes in the opposite direction, however it is only
operative for clusters with short relaxation times. As clusters are possibly
more massive than estimated from observations but also less concentrated
(lower King fitting parameter) it is not clear how tides and other disruptive
effects will shape up the CMF, especially if the host galaxy itself is out of
equilibrium.  A set of fully three-dimensional N-body simulations could
enhance our knowledge of the influence of mass segregation on longer
time-scales and with strong tidal fields found for example in merging
galaxies.

Our models of isolated clusters suffer a few important limitations. We
have mentioned the role that tidal fields will play in removing weakly
bound stars. 
 Another aspect of the problem is the possibly low star formation efficiency
when the cluster forms. We mentioned how gas from stellar winds might impact
on the dynamics (\S\ref{long:evolution}). Residual gas from the formation 
epoch will
also drive much evolution in the early stages by bringing the cluster out of
virial equilibrium (Elson et al. 1989, Kroupa \& Boily 2002, see also Bastian
\& Goodwin 2006). 
 Yet another aspect is our tacit assumption that stars are all
born at the same time and all evolve in unisson. Stars in massive clusters
may well have ages that vary by a few million years. This will have some
bearing on the rise of $\eta$ in the early stages because not all the
stars become remnant at the same time, leading to enhanced mass
segregation and further increase in $\eta$. For instance, the knee seen at
$t \approx 10\U{Myr}$ may well increase to yet higher values before shifting
over to the slower rate of increase that we have advertised 
(Fig.~\ref{max:mass}).

A more severe limitation however, one that will impact on $\eta$ at all times,
is the fraction of primordial binaries.  Tight binaries will survive for eons
and in particular a very large fraction of them will survive for the short
times that are of interest here. The presence of binaries and multiple stars
naturally enhances the observed velocity dispersion which biases the mass 
estimate to larger values through the virial theorem.
However, binaries also instantly broaden
the width of the effective stellar IMF, since, roughly speaking, they will
dynamically act as single stars of mass equal to the sum of their member
stars. If the fraction of primordial binaries is low, the mean stellar mass
will remain unchanged but the maximum mass will effectively double. The net
effect, then, is similar to halving the mass segregation time-scale $t\e{ms}$
by reducing the mean stellar mass. This can be accomplished by increasing the
power index $\alpha$ of the IMF in Eq.~\eqref{eq:IMF}. We have found after 
$\approx
10\U{Myr}$ of evolution for the extreme case where $\alpha = 2.3$ (Salpeter
value) that $\eta$ has nearly increased by $30$ to $50$ per cent in comparison
with the result for the standard case $\alpha = 1.3$.  These considerations
clearly point to yet more rapid evolution and the need for more realistic
models than is affordable here to pin down more precisely the dynamics of
young massive clusters.

\acknowledgments

We thank Douglas Heggie, Onno Pols and Simon Portegies Zwart for discussions. 
We
 also thank Douglas Heggie for providing the motivation for this investigation
during a visit to Strasbourg in 2004, and him and Richard de Grijs for
detailed comments on a draft version of this paper.

{

}

\label{lastpage}

\end{document}